\documentclass[a4paper,11pt]{article}
\usepackage{jheppub}
\usepackage{slashed}
\usepackage{color}
\usepackage{rotating}
\usepackage{cancel}

\newcommand{\gev}{\textrm{GeV}}
\newcommand{\fb}{\textrm{fb}}

\newcommand{\miss}{\textrm{miss}}

\title{\boldmath Higher-Order QCD Predictions for Dark Matter Production in Mono-$Z$ Searches at the LHC}

\author[a,b]{Matthias Neubert,}
\author[a]{Jian Wang}
\author[c]{and Cen Zhang}

\affiliation[a]{PRISMA Cluster of Excellence $\&$ Mainz Institute for Theoretical Physics,
Johannes Gutenberg University, D-55099 Mainz, Germany }
\affiliation[b]{Department of Physics, LEPP, Cornell University, Ithaca, NY 14853, U.S.A.}
\affiliation[c]{Department of Physics, Brookhaven National Laboratory, Upton, N.Y., 11973, U.S.A.}

\emailAdd{neubertm@uni-mainz.de}
\emailAdd{jian.wang@uni-mainz.de}
\emailAdd{cenzhang@bnl.gov}

\abstract{We present theoretical predictions for mono-$Z$ production in the search for dark matter in Run-II at the LHC, including next-to-leading order QCD corrections and parton-shower effects. We consider generic simplified models with vector and scalar $s$-channel mediators. The calculation is performed by implementing the simplified models in the {\sc FeynRules/MadGraph5\_aMC@NLO} framework, which allows us to include higher-order QCD corrections and parton-shower effects in an automated way. We find that these corrections are sizeable and help to reduce the theoretical uncertainties. We also investigate the discovery potential in several benchmark scenarios in the 13 TeV run at the LHC.}

\arxivnumber{1509.05785}
\preprint{MITP/15-069}

\begin{document}
\maketitle
\flushbottom

\section{Introduction}
\label{sec:introduction}

The nature of dark matter (DM) is still a mystery in our knowledge about the universe. In spite of increasingly accurate cosmological and astrophysical observations confirming the existence of cold DM (see e.g.\ \cite{Baer:2014eja,Gelmini:2015zpa} for recent reviews), its properties, such as spin, mass and flavor structure, remain elusive. The most often considered theoretical candidate for cold DM is a weakly-interacting massive particle (WIMP), which may be detected via non-gravitational interactions. However, there are no candidates for WIMPs in the standard model (SM) of particle
physics. Therefore, any theory trying to describe DM requires an extension of the SM.

Searches for DM have been carried out in many ways, including cosmology and astrophysics observations, direct- and indirect-detection experiments, and searches at particle colliders. Recently there has been a rise in the interest in collider searches for DM, as the LHC opens up new energy regimes and more search possibilities, and its sensitivity continues to be improved as the integrated luminosity increases. The generic DM signal at the LHC consists of the production of one or more SM particles accompanied by missing energy. Using the data accumulated in the first stage of running, the ATLAS and CMS Collaborations have already searched for DM in events with missing transverse momentum associated with a single particle $X$, i.e.\ a mono-$X$ signal, where $X$ may be a jet, a photon, or a $W$ or $Z$ boson \cite{ATLAS:2012zim,Aad:2012fw,Aad:2013oja,Aad:2014vka,Khachatryan:2014rra,Khachatryan:2014rwa}. So far no excess has been discovered, and limits have been set on the energy scale of the DM interaction and the mass of the DM particles.

Stringent constraints on spin-independent and spin-dependent DM interactions have already been obtained, respectively, by the LUX experiment \cite{Akerib:2013tjd} and by mono-$X$ searches at the LHC \cite{ATLAS:2012zim,Aad:2012fw,Aad:2013oja,Aad:2014vka,Khachatryan:2014rra,Khachatryan:2014rwa}. These limits have been derived in the framework of effective field theory (EFT) \cite{Cao:2009uw,Bai:2010hh,Goodman:2010ku,Beltran:2010ww,Fox:2011pm}, which has the advantage of being model independent and involving only a few parameters, thus allowing for easy comparison between theoretical predictions and various kinds of experiments. However, the EFT is only a low-energy approximation of more complete models. Its validity should be carefully checked when applied to LHC processes. The issue has been widely discussed in the literature, see e.g.\ \cite{Bai:2010hh,Goodman:2010ku,Fox:2011pm,Shoemaker:2011vi,Busoni:2013lha,Buchmueller:2013dya,Busoni:2014sya,Busoni:2014haa}. A potential problem is that the requirement that the fundamental energy scale of the EFT be much larger than the momentum transferred to the DM particles is not always satisfied over the entire region of phase space probed by the experimental searches at the LHC. For example, a mediator with mass in the few-TeV range, which one would integrate out in the EFT approach, could become accessible in events with very high $p_T$, thus rendering the EFT assumption invalid. This issue will become more pressing in Run-II of the LHC. As a result, an alternative framework based on so-called simplified models has been proposed as the standard for future DM searches \cite{Alwall:2008ag,Alves:2011wf,Goodman:2011jq,An:2012va,Frandsen:2012rk,An:2013xka,DiFranzo:2013vra,Papucci:2014iwa,Berlin:2014tja,Buchmueller:2014yoa,
Abdallah:2014hon,Malik:2014ggr,Buckley:2014fba,Harris:2014hga,Alves:2015pea,Jacques:2015zha,Haisch:2015ioa,Alves:2015mua,Harris:2015kda} (see \cite{Abdallah:2015ter,Abercrombie:2015wmb} for recent reviews). The idea behind simplified models is to incorporate the most relevant degrees of freedom of the underlying (UV complete) model at accessible energies, while keeping much of the simplicity of EFT. Moreover, they may provide new search signals besides the missing-energy signature, such as the direct production of the mediators. Meanwhile, simplified models also add some complexity. Unlike in EFT, the couplings between the mediator and the DM and SM particles need to be specified, and the width of the mediator, which enters as a new parameter, needs to be calculated. As a consequence, the results can depend on the couplings in a rather nontrivial way.

Current theoretical predictions for DM production at the LHC have mostly been provided at leading order (LO) in QCD, although some next-to-leading order (NLO) QCD results exist within the EFT framework \cite{Wang:2011sx,Huang:2012hs,Fox:2012ru,Haisch:2012kf,Haisch:2013ata,Mao:2014rga}. With the data of the 13 TeV collision, it is expected that more precise constraints can be obtained. In order to match the experimental accuracy, it is important to have complete theoretical predictions with NLO accuracy, including parton-shower effects. Such higher-order corrections tend to increase the cross sections by a few tens of percent and reduce the scale uncertainties, thus making the theoretical predictions more reliable. In addition, with more final state particles, kinematic distributions can be predicted more reliably at NLO, serving as a reference for experimentalists to set appropriate kinematic cuts.

In this paper we provide theoretical predictions for mono-$Z$ production at the LHC including NLO QCD corrections and the parton shower. To the best of our knowledge, such predictions have not been presented before. The mono-$Z$ channel is particularly important for DM searches at the LHC. Previous studies of this channel within the EFT framework have been based on LO predictions \cite{Carpenter:2012rg,Bell:2012rg,Chen:2013gya,Alves:2015dya,Crivellin:2015wva}. The ATLAS Collaboration has reported constraints on the $pp\to Z+\cancel{E}_T$ cross section, where the $Z$ boson is reconstructed either via a leptonic decay \cite{Aad:2012awa} or a hadronic decay \cite{Aad:2013oja}. Our results provide a more solid theoretical basis for future studies in this channel.

\begin{figure}\centering
\includegraphics[width=0.37\linewidth]{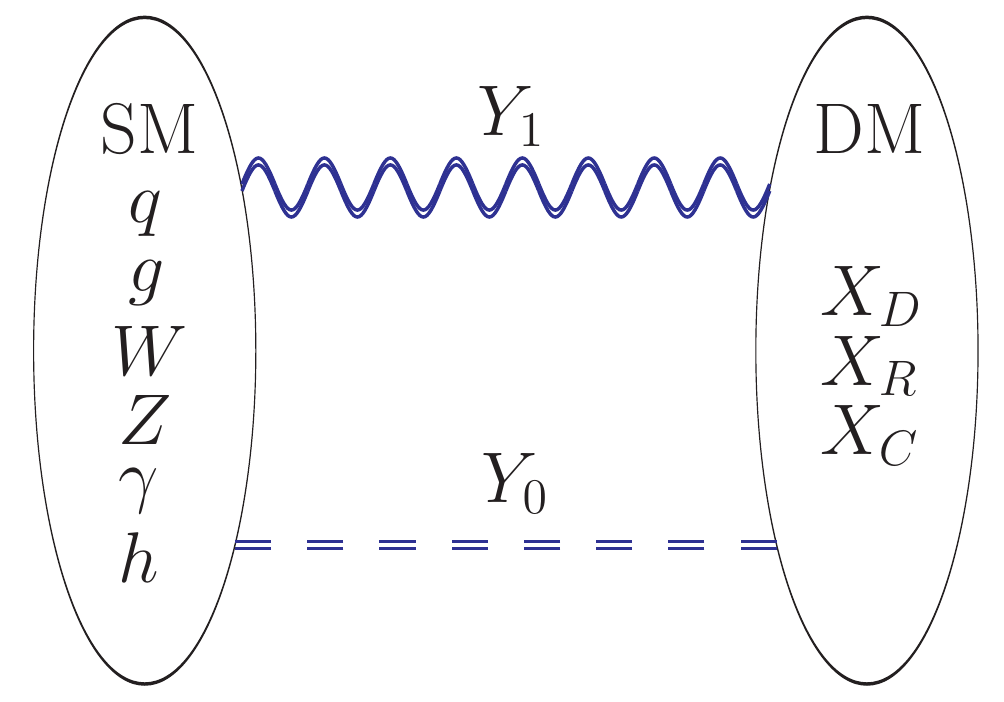}
\caption{Illustration of the connection between SM and DM particles via messengers $Y_i$.}
\label{fig:dmsm}
\end{figure}

We emphasize that our work is part of a larger project aiming at providing complete NLO predictions for all relevant DM search channels at the LHC based on the automated framework of {\sc FeynRules} \cite{Alloul:2013bka} and {\sc MadGraph5\_aMC@NLO} \cite{Alwall:2014hca}. As a first step, we have implemented the simplified models with spin-0 and spin-1 $s$-channel mediators, shown in figure~\ref{fig:dmsm}. This work has been coordinated with the authors of \cite{Backovic:2015soa}; however, compared to this reference we add some specific interactions and corresponding higher-dimensional  operators contributing to the $V+\cancel{E}_T$ production channels (with $V=Z,W,\gamma$) so as to allow for more complete studies of mono-$V$ signatures. There are two main features of this approach. First, it realizes full automatization from drawing Feynman diagrams to generating collision events at NLO matched with parton shower and is thus a convenient tool for both theorists and experimentalists. Second, with a few extra lines of code one can in principle calculate \textit{any} mono-$X$ signal, and hence complete NLO results for DM production become readily available. In this work, we focus on the mono-$Z$ production to illustrate the capabilities of such an approach. Throughout this paper we stick closely to the benchmark points suggested by the ATLAS/CMS DM forum \cite{Abercrombie:2015wmb}, but we stress that our implementation is far more general and different scenarios can be studied as well. An analogous application of this framework to the mono-jet channel has recently been discussed in \cite{Backovic:2015soa}. In addition, some loop-induced DM production modes have been explored in \cite{Mattelaer:2015haa} based on recent tools for calculating loop-level processes \cite{Hirschi:2015iia}. The analysis of other search channels is left for future work.

This paper is organized as follows. We first introduce the simplified DM models and effective operators in section~\ref{sec:model} and describe their implementation in section~\ref{sec:mg}. We then present QCD NLO results for total cross sections in section~\ref{sec:tot} and kinematic distributions in section~\ref{sec:kin}. The discovery potential of mono-$Z$ signals at the 13 TeV LHC is discussed in section \ref{sec:lhc}. Section~\ref{sec:Conclusion} contains some conclusions and outlook.

\section{Simplified models for dark matter production via $s$-channel mediators}
\label{sec:model}

We start by defining the simplified models underlying our analysis. We consider three types of DM particles $X_i$, namely a Dirac fermion $X_D$, a real scalar $X_R$, and a complex scalar $X_C$. We also consider two kinds of mediators $Y_i$, namely a real scalar $Y_0$ and a vector $Y_1$, see figure~\ref{fig:dmsm}. These particles are assumed to be singlets under the SM gauge group. Our notations for particles and coupling constants are consistent with those employed in \cite{Backovic:2015soa}, but we explore a larger number of interactions in the present work. Model files, which specify our simplified models, have been released on the website \cite{dmsimp}.

We consider different types of interactions between the SM, the dark sector and the mediators. The renormalizable interactions of the messengers with the DM are described by the Lagrangians
\begin{equation}\label{LDM}
\begin{aligned}
   {\cal L}_{\rm DM}^{Y_1}
   &= \frac{i}{2}\,g^{V}_{X_C} \big[ X_C^* (\partial_{\mu} X_C)
    - (\partial_{\mu} X_C^*) X_C \big] Y_1^{\mu}
   + \overline{X}_D \gamma_{\mu} \big( g^{V}_{X_D}
   + g^{A}_{X_D}\gamma_5 \big) X_D\,Y_1^{\mu} \,, \\
   {\cal L}_{\rm DM}^{Y_0}
   &= \frac{1}{2} M_{X_R}\,g^{S}_{X_R} X_R X_R Y_0 + M_{X_C}\,g^{S}_{X_C} X_C^* X_C Y_0
    + \overline X_D \big( g^{S}_{X_D}+ig^{P}_{X_D}\gamma_5 \big) X_D\,Y_0 \,.
\end{aligned}
\end{equation}
We omit quadratic terms in the messenger fields, which will play no role in our analysis. The leading interactions of the scalar messenger $Y_0$ with scalar DM fields are described by dimension-3 operators. We use the DM masses to serve as natural scales in these terms, but one is always free to readjust the coupling strengths if a different normalization scale is employed. The renormalizable interactions of the messenger fields with SM quarks are described by the Lagrangians
\begin{equation}\label{LSM}
\begin{aligned}
   {\cal L}_{\rm SM}^{Y_1}
   &= \sum_{i,j} \left[ \bar d_i\gamma_{\mu} \big( g^{V}_{d_{ij}} + g^{A}_{d_{ij}}\gamma_5
    \big) d_j + \bar u_i\gamma_{\mu} \big( g^{V}_{u_{ij}} + g^{A}_{u_{ij}}\gamma_5 \big) u_j
    \right] Y_1^{\mu} \,, \\
   {\cal L}_{\rm SM}^{Y_0}
   &= \sum_{i,j} \left[ \bar d_i\,\frac{y_{i}^d}{\sqrt{2}}
    \big( g^{S}_{d_{ij}} + ig^{P}_{d_{ij}}\gamma_5 \big) d_j
    + \bar u_i\,\frac{y_{i}^u}{\sqrt{2}}
    \big( g^{S}_{u_{ij}} + ig^{P}_{u_{ij}}\gamma_5 \big) u_j \right] Y_0 \,,
\end{aligned}
\end{equation}
which refer to the mass basis of the SM quarks. In order to avoid large flavor-changing neutral current interactions, we will assume that the coefficients $g_{q_{ij}}^B$ are flavor diagonal, i.e.\ $g_{q_{ij}}^B=\delta_{ij}\,g_{q_i}^B$ for $B=V,A,S,P$. We also assume that the (pseudo-)scalar couplings are aligned with the SM Yukawa interactions, and hence in the mass basis they are proportional to the Yukawa couplings $y_i^q$ of the various quarks. It then follows that the interactions of the scalar mediator $Y_0$ with light quarks are strongly suppressed. As a consequence, this type of coupling is best probed in the search channel $t\bar t+\cancel{E}_T$ \cite{Cheung:2010zf,Lin:2013sca, Backovic:2015soa}. We emphasize that the operators containing $Y_0$ in (\ref{LSM}) are gauge invariant only after electroweak symmetry breaking, which implies that the coefficients $g_{q_{ij}}^{S,P}$ must implicitly contain a factor $v/\Lambda$ ($v$ is the vacuum expectation value of the Higgs field and $\Lambda$ is a new physics scale),  suppressing these interactions further even for the case of the top quark.

The Lagrangians (\ref{LDM}) and (\ref{LSM}) have been discussed previously in \cite{Backovic:2015soa}; we show them here only for completeness. In addition, we also consider couplings of the mediators with the electroweak bosons $Z$, $W$ and $h$ of the SM. For the $Y_1$ mediator, the corresponding renormalizable couplings are derived from the Lagrangian\begin{align}\label{eq:zmixing}
   {\cal L}_{{\textrm{EW}}}^{Y_1} = g^V_{h}\,\frac{i}{2} \big[ \phi^\dagger (D_\mu\phi)
    - (D_\mu\phi)^\dagger \phi \big] Y_1^\mu \,,
\end{align}
where $\phi$ is the Higgs doublet. A kinetic-mixing term with the hypercharge field of the form $B_{\mu\nu}\,\partial^\mu Y_1^\nu$ can be reduced to the operators contained in ${\cal L}_{\rm SM}^{Y_1}$ in (\ref{LSM}) and $ {\cal L}_{{\textrm{EW}}}^{Y_1}$ in (\ref{eq:zmixing})  using an integration by parts. The interaction in (\ref{eq:zmixing}) can  induce a mixing between $Y_1$ and $Z$, and as a result the couplings between the $Z$ boson and other SM particles are modified. This mixing is constrained by precision electroweak measurements at the permille level \cite{Agashe:2014kda}, which implies the bound (with $g_W$ the SU(2) gauge coupling)
\begin{align}
   \frac{\cos\theta_W\,g^V_h}{g_W}\frac{m_Z^2}{M_{Y_1}^2-m_Z^2}\lesssim 10^{-3} \,.
\end{align}
Note that $g_h^V$ is not necessarily suppressed, since $M_{Y_1}$ can be much larger than $m_Z$. Besides this mixing effect, the Lagrangian (\ref{eq:zmixing}) does not play a role in our analysis of mono-$Z$ production. For the $Y_0$ mediator, the leading interaction terms come from
\begin{align}\label{eq:hmixing}
   {\cal L}_{{\textrm{EW}}}^{Y_0}
   = M_{Y_0}\,g_{h1}^S \left| \phi \right|^2 Y_0 + g_{h2}^S \left| \phi \right|^2 Y_0^2 \,.
\end{align}
When one of the Higgs fields takes its vacuum expectation value, these interactions induce a mixing between $Y_0$ and the SM Higgs field $h$, which causes a universal rescaling of all Higgs couplings. Current measurements of the Higgs couplings imply the bound \cite{Khachatryan:2014jba}
\begin{align}
   \frac{v M_{Y_0}\,g_{h1}^S}{M_{Y_0}^2-m_h^2} < 0.37 \,.
\end{align}
The interactions (\ref{eq:hmixing}) also give rise to Higgs-boson couplings to DM, which have been extensively investigated in the context of Higgs-portal models \cite{Baek:2011aa,LopezHonorez:2012kv,Baek:2012se,Duch:2015jta}. Since our focus here is on the mono-$Z$ signal, we do not take these kind of interactions into account.

\begin{figure}\centering
\includegraphics[width=0.37\linewidth]{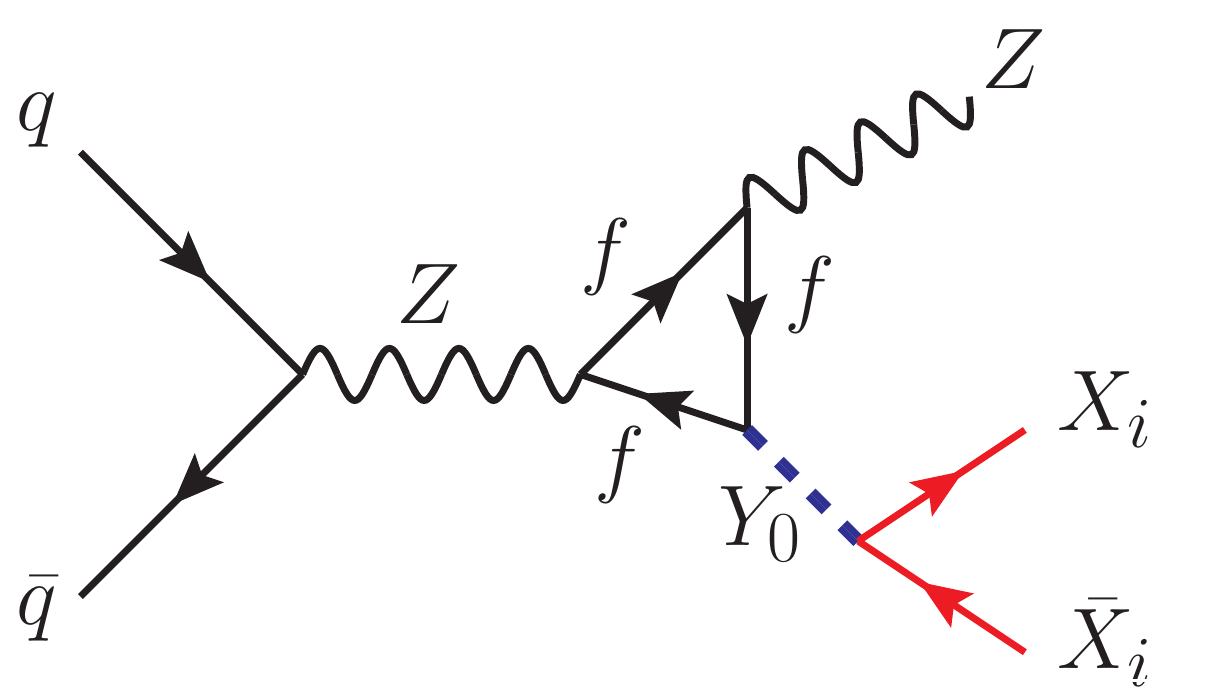}
\caption{Mono-$Z$ production induced by a loop of some heavy fermion $f$.}
\label{fig:monozloop}
\end{figure}

Until now, the interactions between the scalar mediator $Y_0$ and SM particles which we have considered are either strongly suppressed by Yukawa couplings or do not induce a mono-$Z$ signal. This motivates us to include (effective) operators of higher mass dimension. The relevant dimension-5 operators are\footnote{There exist analogous couplings of $Y_0$ to gluons and photons, but they do not contribute to mono-$Z$ production at order $1/\Lambda$. Note, in particular, that the decay chain $gg\to Y_0\to ZZ^*\to Z\nu\bar\nu$, which is an irreducible background to DM searches in the mono-$Z$ channel, comes with a $1/\Lambda^2$ suppression.}
\begin{equation}\label{eq:2.7}
\begin{aligned}
   {\cal L}_{{\textrm{EW,\,dim-5}}}^{Y_0}
   &= \frac{1}{\Lambda}\,\Big[ g^S_{h3} (D^\mu\phi)^\dagger (D_\mu\phi)
    + g^S_B\,B_{\mu\nu} B^{\mu\nu} + g^P_B\,B_{\mu\nu} \tilde B^{\mu\nu} \\
  &\hspace{1.0cm}\mbox{}+ g^S_W W_{\mu\nu}^i W^{i,\mu\nu}
   + g^P_W \,W_{\mu\nu}^i \tilde W^{i,\mu\nu} \Big] Y_0 \,,
\end{aligned}
\end{equation}
where $\tilde V_{\mu\nu}=\frac{1}{2}\epsilon_{\mu\nu\rho\sigma}V^{\rho\sigma}$ are the dual field-strength tensors, and $\Lambda$ is some large energy scale. These effective operators can be induced by loop graphs such as those shown in figure~\ref{fig:monozloop}, where the heavy fermion $f$ is integrated out. The dimension-5 terms in ${\cal L}_{{\textrm{EW,dim-5}}}^{Y_0}$ may be considered as a ``partial'' UV completion of an effective local $VV\bar X_i X_i$ interaction, where the resonance structure of the $s$-channel $Y_0$ mediator can be probed experimentally. The underlying physics is that the SM gauge bosons can couple to the mediator $Y_0$ through loops containing some (very) heavy new particles. Of course, such loop diagrams can also induce operators of higher dimension than~5. Their phenomenological impact will however be very small provided that the new-physics scale $\Lambda$ is sufficiently large.

\begin{figure}\centering
\includegraphics[width=0.9\linewidth]{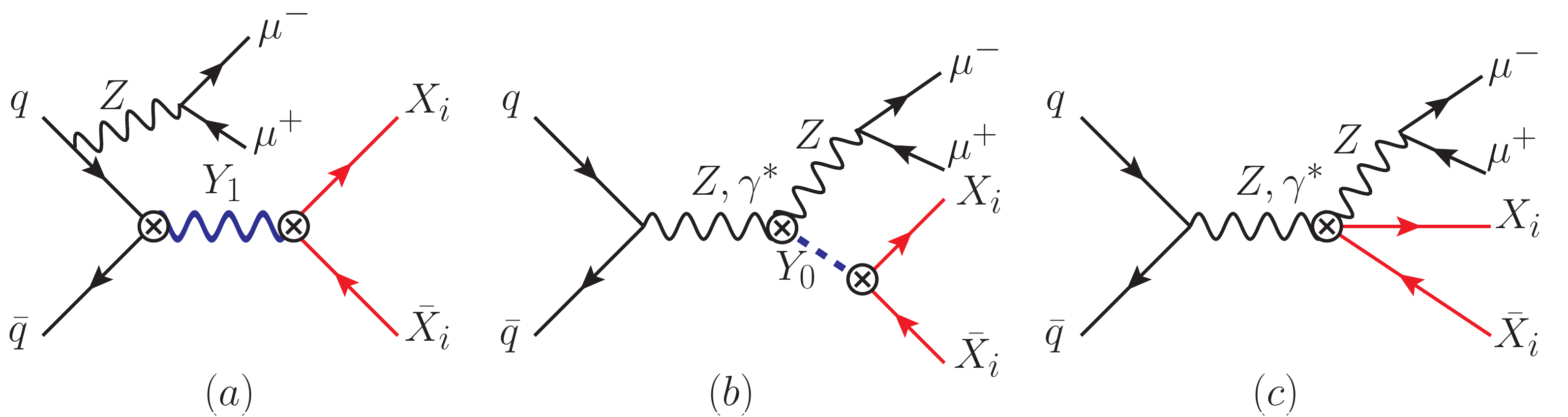}
\caption{Different mono-$Z$ production mechanisms at the LHC. Scenarios $(a)$ and $(b)$ are studied in our analysis
while the scenario $(c)$ is briefly discussed in appendix \ref{app:eft}.}
\label{fig:monoz}
\end{figure}

The above simplified models are fairly general and flexible. It is straightforward to implement them in the universal FeynRules output (UFO) format \cite{Degrande:2011ua}, which allows us to study DM production processes within the {MadGraph5\_aMC@NLO} framework in all relevant production channels, within different benchmark scenarios, and with NLO accuracy. As mentioned earlier, in the present work we focus on the mono-$Z$ signal for illustration purposes, but readers interested in other mono-$X$ signals will find our framework useful as a tool that provides accurate and realistic simulations of various DM production channels. The UFO files for our simplified models are available at the {\sc FeynRules} repository \cite{dmsimp}. Given the list of interaction terms we have specified, the main production mechanisms contributing (at tree level) to the mono-$Z$ signal are shown in the first two graphs in figure~\ref{fig:monoz}. The crossed vertices denote the interactions contained in the above Lagrangians, while all other vertices are those of the SM. In the type-$A$ scenario shown in the first graph, the $Z$ boson is emitted from the initial-state quarks, and the vector mediator $Y_1$ connects the SM to the dark sector. In the type-$B$ scenario indicated by the second diagram, the DM couples to the $Z$ boson through a scalar mediator $Y_0$. In view of the large number of relevant couplings that enter the results, we shall in the following discussion choose a few representative values for illustration purposes. They are listed in table~\ref{tab:coupcomb}. For the masses of DM particles and mediators we follow the suggestions of \cite{Abercrombie:2015wmb}. A grid with the corresponding values is shown in table~\ref{tab:parameter}.

\begin{table}\centering
\begin{tabular}{|c|c|c|c|}
\hline
Scenarios & Dark matter & Relevant couplings & Interactions\\
\hline\hline
\multicolumn{4}{|c|}{Vector mediator $Y_1$} \\
\hline
$A_1$ & $X_D$ & $g^V_q=0.25,\ g^V_{X_D}=1$ & spin-independent \\
$A_2$ & $X_D$ & $g^A_q=0.25,\ g^A_{X_D}=1$ & spin-dependent \\
$A_3$ & $X_C$ & $g^V_q=0.25,\ g^V_{X_C}=1$ & spin-independent \\
\hline
\multicolumn{4}{|c|}{Scalar mediator $Y_0$} \\
\hline
$B_1$ & $X_D$ & $g^S_W=0.25,\ g^S_{X_D}=1,\ \Lambda=3\ \mathrm{TeV}$ & $CP$-even \\
$B_2$ & $X_D$ & $g^P_W=0.25,\ g^P_{X_D}=1,\ \Lambda=3\ \mathrm{TeV}$ & $CP$-odd \\
$B_3$ & $X_C$ & $g^S_{h3}=0.25,\ g^S_{X_C}=1,\ \Lambda=3\ \mathrm{TeV}$ & $CP$-even \\
\hline
\end{tabular}
\caption{Benchmark scenarios with representative values of coupling constants in different DM production scenarios. In each case, the couplings not shown are set to zero.}
\label{tab:coupcomb}
\end{table}

\begin{table}\centering
\begin{tabular}{|c|cccccccccc|}
\hline
$m_{\textrm{DM}}$\,[GeV] & \multicolumn{10}{|c|}{$M_{\textrm{med}}$\,[GeV]} \\
\hline
1 & 10 & 20 & 50 & 100 & 200 & 300 & 500 & 1000 & 2000 & 10000 \\
10 & 10 & 15 & 50 & 100 & & & & & & 10000 \\
50 & 10 & & 50 & 95 & 200 & 300 & & & & 10000 \\
150 & 10 & & & & 200 & 295 & 500 & 1000 & & 10000 \\
500 & 10 & & & & & & 500 & 995 & 2000 & 10000 \\
1000 & 10 & & & & & & & 1000 & 1995 & 10000 \\
\hline
\end{tabular}
\caption{DM masses $m_{\textrm{DM}}$ and mediator masses $M_{\textrm{med}}$ used in the benchmark models.}
\label{tab:parameter}
\end{table}

The third diagram in figure~\ref{fig:monoz} shows a topology not covered by our analysis. If the sector that couples the SM to the DM consists of very heavy particles, then the resulting effective interactions at accessible LHC energies can be represented by contact interactions consisting of both visible-sector and dark-sector fields. This gives rise to local, higher-dimensional operators, which have been studied, e.g., in \cite{Cotta:2012nj,Carpenter:2012rg}. This case is briefly discussed in appendix~\ref{app:eft}. In such a scenario our simplified-model framework does not apply, because there is no mediator involved. In this case an EFT approach can be useful. Indeed, these higher-dimensional operators are still recommended for benchmark studies of mono-$V$ signals in \cite{Abercrombie:2015wmb}, because the UV completions of such interactions have not yet been fully
investigated.

\begin{figure}\centering
\includegraphics[width=0.5\linewidth]{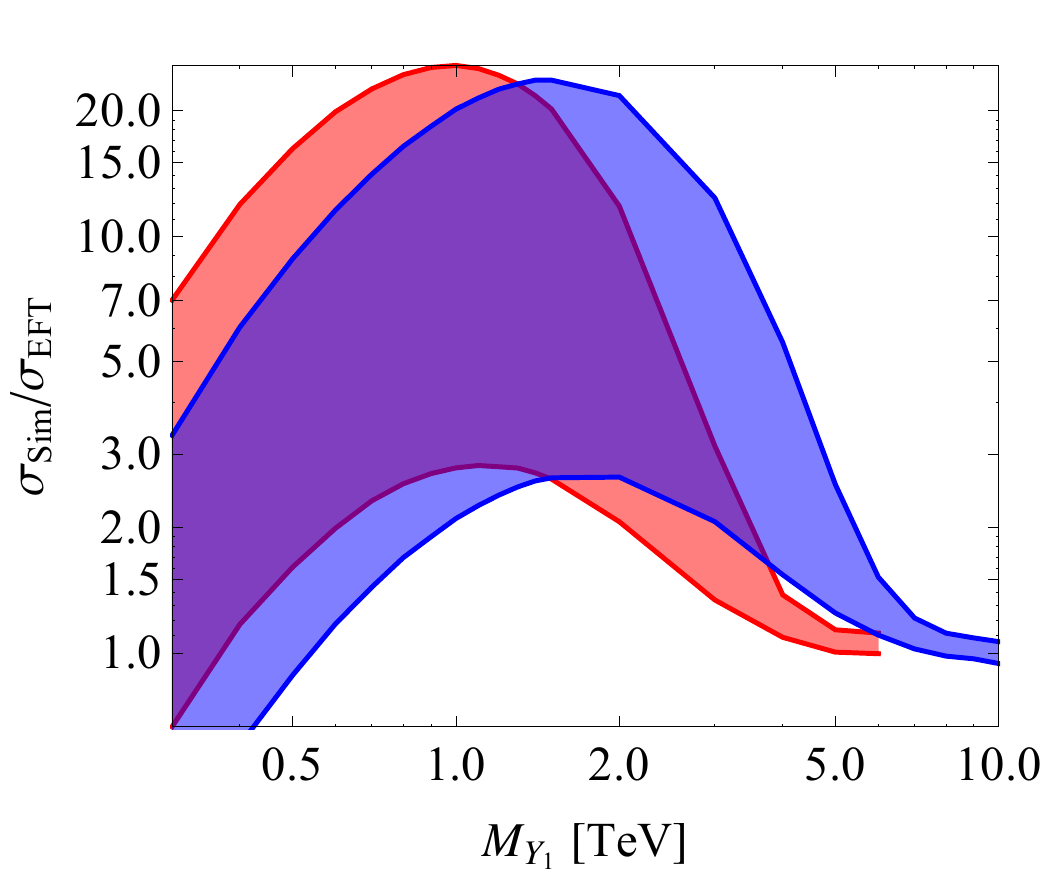}
\caption{Ratio of the mono-$Z$ production cross sections at the 8 TeV (red) and 13 TeV (blue) LHC, computed in simplified models and EFT. The assumed mediator is a vector $Y_1^\mu$. The upper and lower limits of the bands correspond to the cases where $\Gamma_{Y_1}=M_{Y_1}/(8\pi)$ and $\Gamma_{Y_1}=M_{Y_1}/3$, respectively. The DM mass is chosen to be $m_{\textrm{DM}}=100$~GeV.}
\label{eps:com-EFT-monoZ}
\end{figure}

Before closing this section, we would like to give a comparison between the EFT approach and the simplified-model framework for the case of a vector mediator, in order to motivate the use of simplified models in our work. In the limit of a large mediator mass, i.e.\ $M_{Y_i}\gg 2m_{\textrm{DM}}$ and $M_{Y_i}\gg\sqrt{q^2}$, with $q$ the momentum carried by the mediator, the mediator fields can be safely integrated out from the effective theory, resulting in higher-dimensional operators coupling the SM and DM particles in EFT. However, this limit is not always applicable at the LHC. To see how exactly the EFT description approaches the simplified models, we compare the total cross sections computed in the two schemes. We take the mono-$Z$ production process induced by a vector mediator as a specific example. We set the DM mass to 100 GeV and all the relevant couplings equal to 1 for simplicity, and we vary the mediator mass between 500 GeV to 10 TeV. In order to investigate the dependence of the cross section on the width of the mediator particle, we consider the two values $\Gamma_{Y_1}=M_{Y_1}/3$, corresponding to a rather broad resonance, and $\Gamma_{Y_1}=M_{Y_1}/(8\pi)$, corresponding to a narrow resonance \cite{Khachatryan:2014rra}. No experimental cuts are applied. The results are presented in figure~\ref{eps:com-EFT-monoZ}, which shows that the cross section calculated in simplified models is generally much larger than that computed in EFT. The enhancement is due to the $s$-channel resonance, and it is larger in the case of a narrow width. This comparison indicates that the EFT does not provide an appropriate framework for predicting DM production at the LHC. In order to capture the effects of $s$-channel resonane, even in cases where the resonance mass lies in the few-TeV range, simplified models should be used.

\section{Implementation and validations}
\label{sec:mg}

Our computations are performed in the framework of {\sc MadGraph5$\_$aMC@NLO} \cite{Alwall:2014hca}. The simplified models are implemented in the UFO format \cite{Degrande:2011ua} by making use of the {\sc FeynRules} package \cite{Alloul:2013bka}. The one-loop corrections are computed using the {\sc MadLoop} program \cite{Hirschi:2011pa}. It is based on the OPP method \cite{Ossola:2006us,Ossola:2007ax}, in which the ultraviolet divergences and rational $R_2$ terms are calculated automatically by means of the {\sc NLOCT} package \cite{Degrande:2014vpa}. The infrared subtraction terms for real emissions are generated by {\sc MadFKS} \cite{Frederix:2009yq}. The matching to the parton shower is performed using the {\sc MC@NLO} framework \cite{Frixione:2002ik}.

We have validated our implementation in several ways. We have calculated the virtual QCD corrections for mono-$Z$ production via the scalar mediator analytically and obtained results that agree with {\sc MadLoop}. We have compared the cross sections for the mono-jet production processes\footnote{Unfortunately, the mono-$Z$ production process has not yet been implemented in {\sc MCFM}.}
induced by vector and axial-vector currents with results from {\sc MCFM} and found good agreement. Finally, by adjusting the couplings and masses of specific mediators so as to mimic SM particles, we can compare certain DM production processes with corresponding SM processes computed using {\sc MadGraph5\_aMC@NLO}. For example, by setting $M_{Y1}=m_Z$ we can compare $Y_1+Z$ production with $ZZ$ production and $Y_1+h$ production with $Z+h$ production, both at LO and NLO. In addition, by setting $M_{Y0}=m_h$ we can compare $Y_0+Z$ production with $Z+h$ production. In all cases we have found that the results agree within the statistic errors.

\section{Total cross sections}
\label{sec:tot}

We start by presenting our predictions for the total cross sections for mono-$Z$ production at the 13 TeV LHC in the benchmark scenarios defined in table~\ref{tab:coupcomb}, assuming that the $Z$ boson decays to $\mu^+\mu^-$. We choose NN23LO1 and NN23NLO PDF sets \cite{Ball:2013hta} for LO and NLO results and use dynamical factorization and renormalization scales, with the default values $\mu_f=\mu_r=\frac{1}{2}H_T=\frac{1}{2}\sum_i\sqrt{p_{T,i}^2+m_i^2}$, where the sum is over all final-state particles. The DM and mediator masses are varied over the values shown in table~\ref{tab:parameter}. Our parameter choices are in agreement with the DM benchmark models suggested in \cite{Abercrombie:2015wmb} for the early phase of the LHC Run-II. The width of the mediator $\Gamma_{\textrm{med}}$ is determined using the {\sc MadWidth} module \cite{Alwall:2014bza}. It depends on the masses and coupling constants specific to each benchmark scenario, but in general, $\Gamma_{\textrm{med}}/M_{\textrm{med}}$ is about a few percent in our parameter choices.
In this section we do not consider any kinematic cuts. Their effects will be discussed in the next section.

\subsection{Benchmark scenarios A$_i$}

We first consider the benchmark scenarios $A_1$, $A_2$ and $A_3$, which correspond to simplified models with an $s$-channel spin-1 mediator $Y_1$. The scenario $A_2$ is perhaps the most interesting one, since it involves spin-dependent couplings to the DM, for which the LHC can set more stringent bounds than direct-detection experiments. The total cross sections and $K$-factors in benchmark scenario scenario $A_2$ are shown in table~\ref{tab:nlo2} and figure~\ref{fig:dmmed_a2}, while those for scenarios $A_1$ and $A_3$ are given in appendix~\ref{app:results}.

\begin{table}
\small\centering
\begin{tabular}{|c|cccccccccc|}
\hline
 & \multicolumn{10}{|c|}{$M_{\textrm{med}}$\,[GeV]} \\
\cline{2-11}
 & 10 & 20 & 50 & 100 & 200 & 300 & 500 & 1000 & 2000 & 10000 \\
$m_{\textrm{DM}}$\,[GeV] & & (15) & & (95) & & (295) & & (995) & (1995) & \\
\hline
1 & 8.5 & 3.5 & 1.0 & 0.35 & 0.10 & 4.5e-2 & 1.3e-2 & 1.7e-3 & 1.1e-4 & 1.3e-8 \\
10 & 4.6e-2 & 5.8e-2 & 0.90 & 0.34 & & & & & & 1.3e-8 \\
50 & 2.5e-3 & & 2.9e-3 & 6.6e-3 & 8.0e-2 & 4.1e-2 & & & & 1.2e-8 \\
150 & 2.0e-4 & & & & 3.0e-4 & 8.5e-4 & 8.8e-3 & 1.6e-3 & & 1.0e-8 \\
500 & 3.5e-6 & & & & & & 4.5e-6  & 2.8e-5 & 7.8e-5 & 4.1e-9 \\
1000 & 1e-7 & & & & & & & 1.4e-7 & 1.3e-6 & 9.4e-10 \\
\hline
$m_{\textrm{DM}}$\,[GeV]& \multicolumn{10}{|c|}{$K$-factor} \\
\hline
1 & 1.57 & 1.46 & 1.49 & 1.48 & 1.42 & 1.39 & 1.38 & 1.35 & 1.29 & 1.29 \\
10 & 1.49 & 1.50 & 1.48 & 1.47 & & & & & & 1.29 \\
50 & 1.41 & & 1.42 & 1.43 & 1.42 & 1.41 & & & & 1.29 \\
150 & 1.38 & & & & 1.38 & 1.39 & 1.40 & 1.36 & & 1.29 \\
500 & 1.33 & & & & & & 1.34 & 1.36 & 1.29 & 1.23 \\
1000 & 1.21 & & & & & & & 1.22 & 1.27 & 1.09 \\
\hline
\end{tabular}
\caption{NLO total cross sections (in pb) and $K$-factors for mono-$Z$ production in the channel $pp\to Z(\to\mu^+\mu^-)+\cancel{E}_T$ at the 13 TeV LHC in benchmark scenario $A_2$. We use the short-hand notation ``e-$n$'' for $10^{-n}$. The values of $M_{\textrm{med}}$ shown in parenthesis are used at threshold $M_{\textrm{med}}=2m_{\textrm{DM}}$.}
\label{tab:nlo2}
\end{table}

We observe that, for fixed mass of the mediator, the cross section is rather insensitive to the DM mass as long as it is below the threshold $m_{\textrm{DM}}=M_{\textrm{med}}/2$, but decreases rapidly for heavier DM masses. For a fixed DM mass, the cross section increases when $M_{\textrm{med}}<2m_{\textrm{DM}}$ and decreases when $M_{\textrm{med}}>2m_{\textrm{DM}}$. These are typical features for $s$-channel resonance production, which are not captured in an EFT approach. From appendix~\ref{app:results}, we also observe that the cross sections for benchmark scenario $A_3$ are approximately one order of magnitude smaller than those in scenarios $A_1$ and $A_2$. This suggests that it may be more promising to search for fermionic DM than for scalar DM. The impact of QCD corrections on the cross sections is reflected by the $K$-factors, which are defined as the ratios of the cross sections computed at NLO and LO. We find that the $K$-factors are nearly the same for all three scenarios, indicating that the QCD corrections are rather insensitive to the coupling structure of the simplified model. The $K$-factors decrease as $M_{\textrm{med}}$ and $m_{\textrm{DM}}$ increase, but mostly stay in a range of roughly $1.3-1.5$. We thus conclude that NLO QCD corrections have a noticeable impact on the
mono-$Z$ signal.

\begin{figure}\centering
\includegraphics[width=0.5\linewidth]{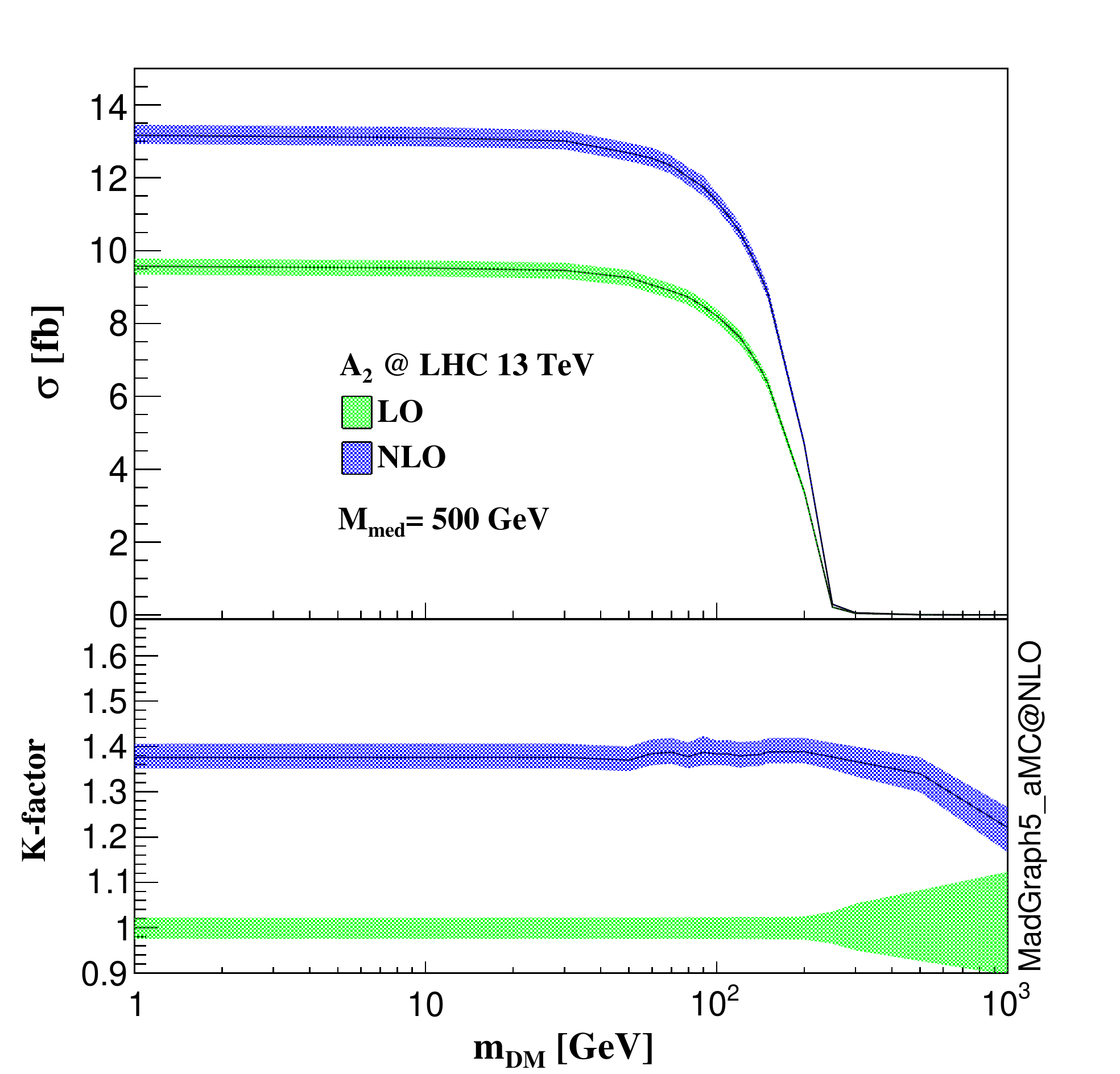}\includegraphics[width=0.5\linewidth]{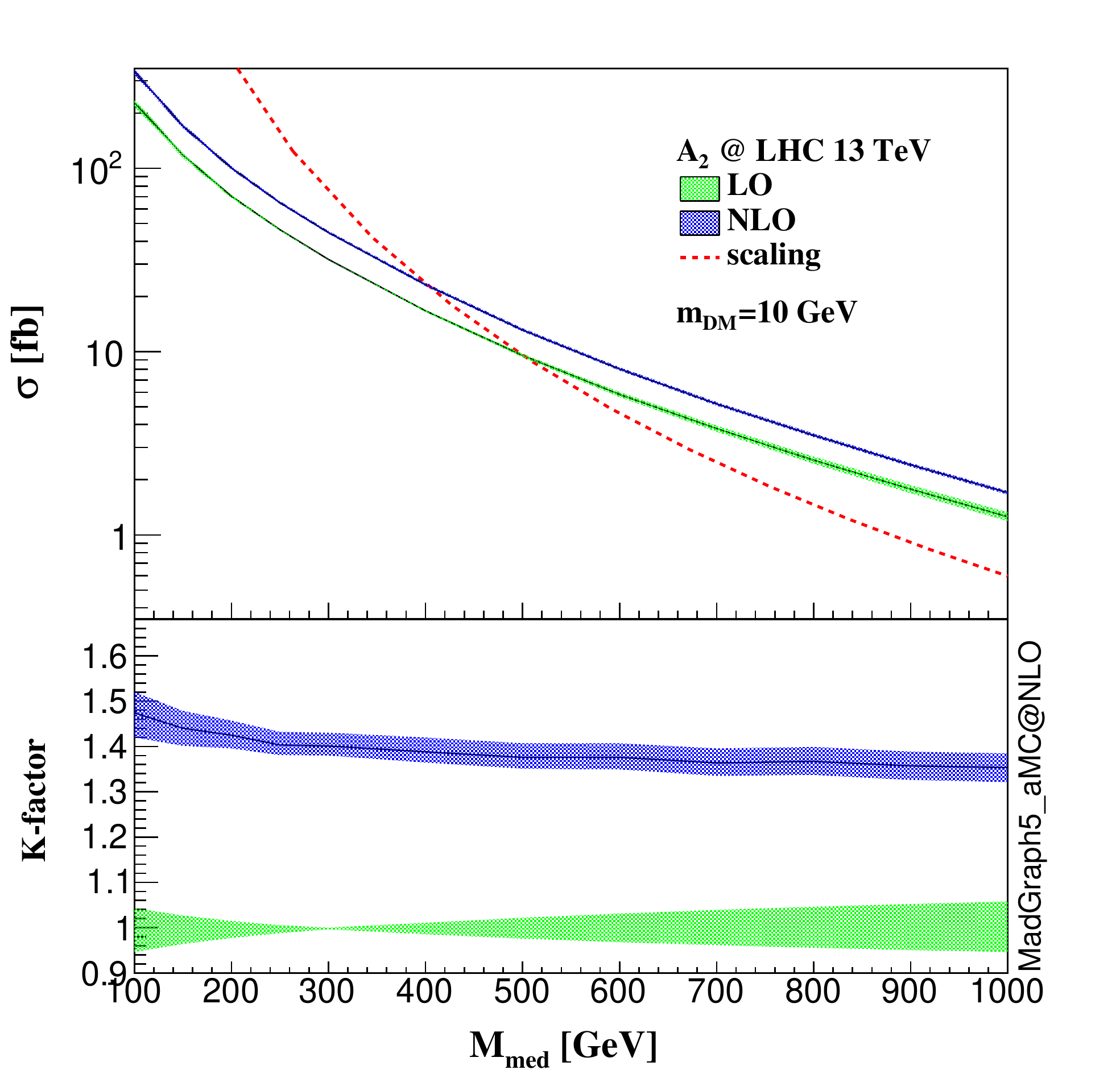}
\caption{Benchmark scenario $A_2$: The $pp\to Z(\to\mu^+\mu^-)+\cancel{E}_T$ cross section and $K$-factor as functions of the DM mass (left) and the mediator mass (right). The dashed red line in the right plot shows a simple scaling behavior $\sigma\propto M_{\rm mad}^{-4}$ predicted by the EFT approach upon integrating out the mediator. The  bands represent the scale uncertainties, as estimated by varying the scales $\mu_f$ and $\mu_r$ independently by a factor~2 about their default values.}
\label{fig:dmmed_a2}
\end{figure}

The left plot in figure~\ref{fig:dmmed_a2} shows that, for a fixed mediator mass $M_{\textrm{med}}=500$~GeV, the cross section is almost unchanged when $m_{\textrm{DM}}$ increases from 1~GeV to 50~GeV, but starts to decrease when $m_{\textrm{DM}}$ exceeds 50~GeV. When the DM mass exceeds the threshold $M_{\textrm{med}}/2$ for on-shell mediator production the cross section dies out very quickly. It is interesting to compare this feature with the results obtained in \cite{Huang:2012hs,Fox:2012ru} for the similar case of mono-$\gamma$ production\footnote{Prior to our work, there did not exist any NLO results on mono-$Z$ production.}
using an EFT approach, where it was found that the cross section drops beyond $m_{\textrm{DM}}\approx 100$~GeV, but does not become negligibly small until $m_{\textrm{DM}}\approx 1000$~GeV. The reason for the discrepancy is that the invariant mass of the DM pair does not pass a threshold if the process is described using an EFT framework. The right plot in the figure shows that the cross section decreases as the mediator mass increases. However, this decrease is considerably slower than a simple scaling law $\sigma\propto M_{\rm mad}^{-4}$ expected in the EFT context. From both plots, we observe an almost constant $K$-factor at $K\approx 1.4$. Note that the scale uncertainties are not significantly reduced when comparing the LO and NLO predictions, the reason being that the LO prediction is independent of~$\alpha_s$.

\subsection{Benchmark scenarios B$_i$}

\begin{table}[htb]
\small\centering
\begin{tabular}{|c|cccccccccc|}
\hline
 & \multicolumn{10}{|c|}{$M_{\textrm{med}}$\,[GeV]} \\
\cline{2-11}
 & 10 & 20 & 50 & 100 & 200 & 300 & 500 & 1000 & 2000 & 10000 \\
$m_{\textrm{DM}}$\,[GeV] & & (15) & & (95) & & (295) & & (995) & (1995) & \\
\hline
1 & 1.2e-2 & 7.2e-3 & 2.5e-3 & 9.8e-4 & 3.3e-4 & 1.5e-4 & 5.1e-5 & 6.9e-6 & 3.2e-7 & 3.1e-11 \\
10 & 8.8e-5 & 1.1e-4 & 2.5e-3 & 9.7e-4 & & & & & & 3.0e-11 \\
50 & 6.5e-6 & & 7.5e-6 & 1.6e-5 & 3.3e-4 & 1.6e-4 & & & & 3.2e-11 \\
150 & 5.8e-7 & & & & 8.5e-7 & 2.4e-6 & 5.1e-5 & 6.7e-6 & & 2.5e-11 \\
500 & 9.5e-9 & & & & & & 1.3e-8 & 8.8e-8 & 2.8e-7 & 8.6e-12 \\
1000 & 9.7e-9 & & & & & & & 1.1e-7 & 2.9e-7 & 8.7e-12 \\
\hline
$m_{\textrm{DM}}$\,[GeV] & \multicolumn{10}{|c|}{$K$-factor} \\
\hline
1 & 1.43 & 1.42 & 1.39 & 1.35 & 1.33 & 1.30 & 1.32 & 1.31 & 1.17 & 1.21 \\
10 & 1.37 & 1.37 & 1.37 & 1.33 & & & & & & 1.17 \\
50 & 1.32 & & 1.31 & 1.32 & 1.32 & 1.32 & & & & 1.29 \\
150 & 1.31 & & & & 1.32 & 1.32 & 1.31 & 1.28 & & 1.22 \\
500 & 1.23 & & & & & & 1.25 & 1.27 & 1.18 &  1.10 \\
1000 & 1.25 & & & & & & & 1.26 & 1.17 & 1.13 \\
\hline
\end{tabular}
\caption{NLO total cross sections (in pb) and $K$-factors for mono-$Z$ production in the channel $pp\to Z(\to\mu^+\mu^-)+\cancel{E}_T$ at the 13 TeV LHC in benchmark scenario $B_1$. We use the short-hand notation ``e-$n$'' for $10^{-n}$. The values of $M_{\textrm{med}}$ shown in parenthesis are used at threshold $M_{\textrm{med}}=2m_{\textrm{DM}}$.}
\label{tab:nlo4}
\end{table}

We now move to benchmark scenarios $B_1$, $B_2$ and $B_3$, which correspond to simplified models with an $s$-channel spin-0 mediator $Y_0$ coupling the DM to the electroweak sector of the SM. The NLO cross sections and $K$-factors for scenario $B_1$ are shown in table~\ref{tab:nlo4} and figure~\ref{fig:dmmed_b1}, while the corresponding results for scenarios $B_2$ and $B_3$ are given in appendix~\ref{app:results}. We first notice that the cross sections in scenario $B_1$, and even more so in scenarios $B_2$ and $B_3$, are much smaller than those in the three benchmark scenarios $A_i$. One of the obvious reasons is that the relevant operators are of dimension-5 and thus suppressed by the large underlying mass scale, which is chosen to be 3 TeV. One can estimate the total cross sections for lower values of the new-physics scale by multiplying them with $(3~\mbox{TeV}/\Lambda)^2$, even though the actual results will be slightly different as the width of the mediator also depends on $\Lambda$. Next, we notice that benchmark scenario $B_2$ with $CP$-odd interactions has comparable production cross sections to scenario $B_1$ with $CP$-even interactions, while scenario $B_3$ leads to much smaller cross sections. This is because the $W_{\mu\nu}^i W^{i,\mu\nu}$ and $W_{\mu\nu}^i\tilde W^{i,\mu\nu}$ operators in (\ref{eq:2.7}) each give rise to two powers of momentum in the amplitude, which strongly enhances the cross sections, while the $(D^\mu\phi)^\dagger(D_\mu\phi)$ operator in (\ref{eq:2.7}) leads to a constant $v^2$ factor. On the other hand, the $K$-factors do not show significant differences in the three cases, and we again find $K\approx 1.3-1.5$ for most regions of parameter space.

\begin{figure}\centering
\includegraphics[width=0.5\linewidth]{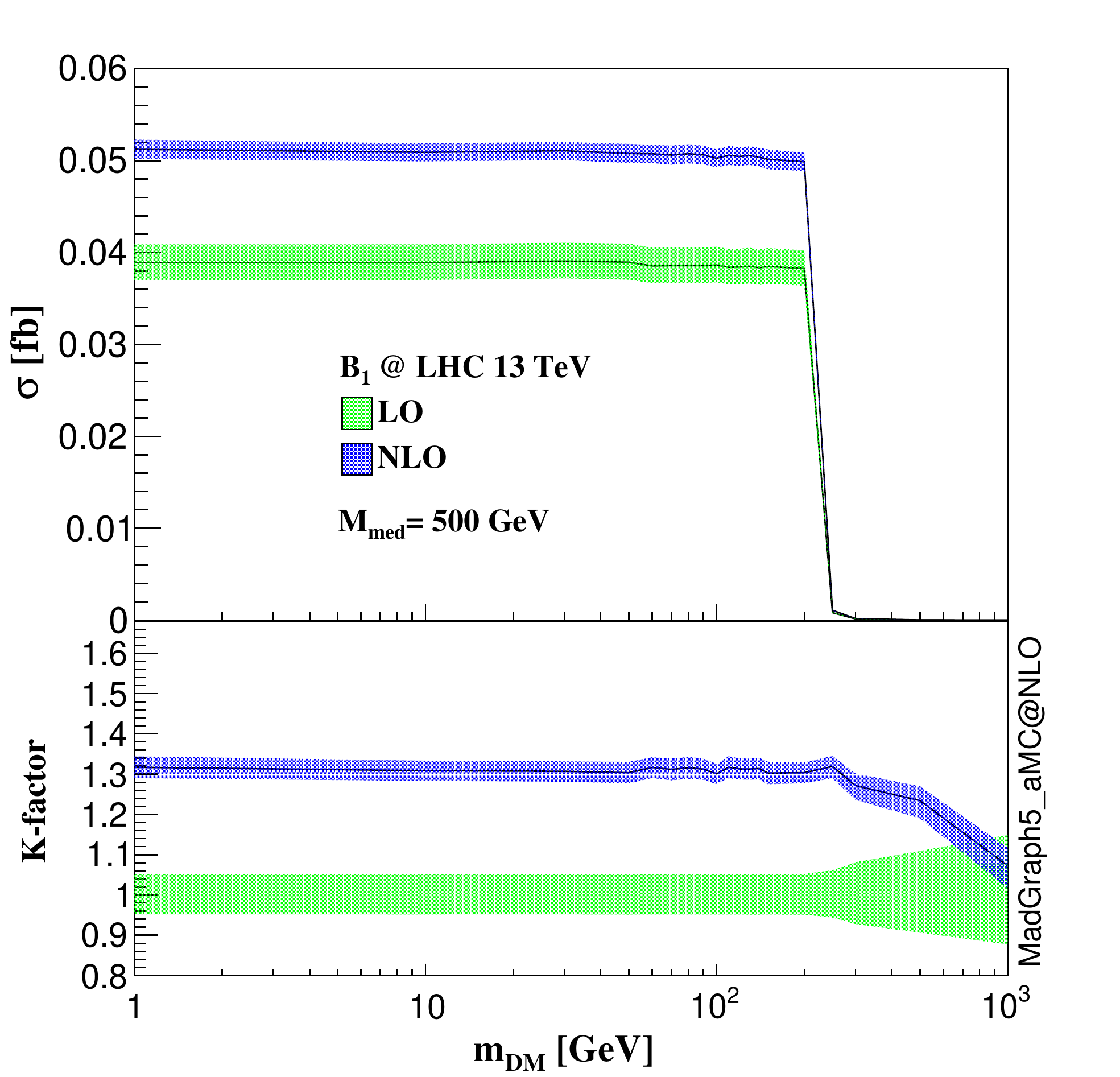}\includegraphics[width=0.5\linewidth]{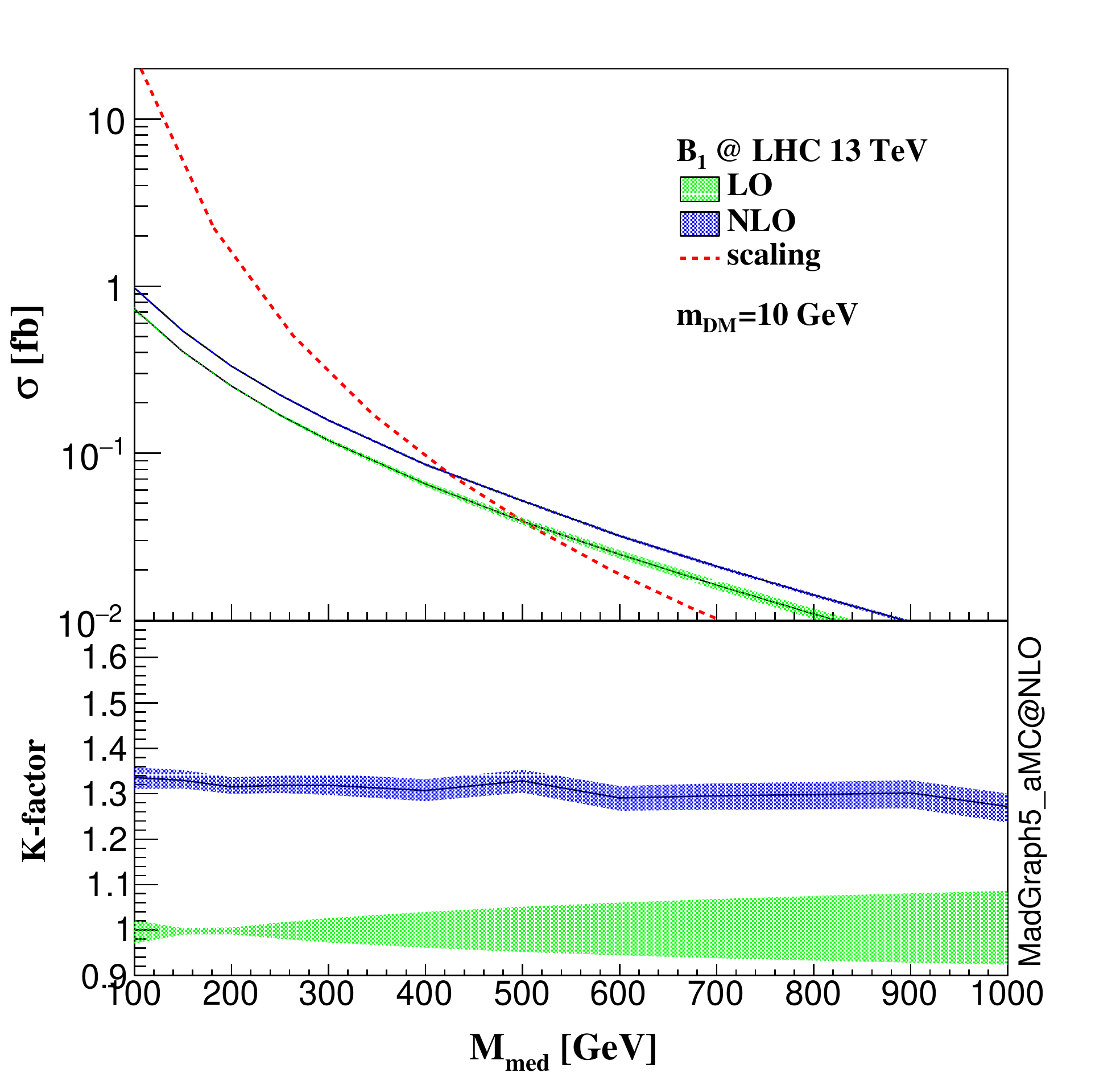}
\caption{Benchmark scenario $B_1$: The $pp\to Z(\to\mu^+\mu^-)+\cancel{E}_T$ cross section and $K$-factor as functions of the DM mass (left) and the mediator mass (right). The dashed red line in the right plot shows a simple scaling behavior $\sigma\propto M_{\rm mad}^{-4}$ predicted by the EFT approach upon integrating out the mediator. The  bands represent the scale uncertainties, as estimated by varying the scales $\mu_f$ and $\mu_r$ independently by a factor~2 about their default values.}
\label{fig:dmmed_b1}
\end{figure}

Figure~\ref{fig:dmmed_b1} shows the cross section in benchmark scenario $B_1$ as a function of the DM (left) and mediator (right) masses. The behavior is quite similar to that observed for scenario $A_2$ in figure~\ref{fig:dmmed_a2}, but the drop-off near the threshold seen in the left plot is much more steep in the case of $B_1$. Compared with benchmark scenario $A_{2}$, the $K$-factor is slightly smaller in the present case, while the reduction of the scale uncertainties at NLO is more pronounced.

\section{Kinematic distributions}
\label{sec:kin}

The total cross sections presented above display some of the basic features of the mono-$Z$ production process at the LHC. However, for practical measurements kinematic cuts always play a crucial role. The inclusion of NLO QCD corrections and parton-shower effects provide us with a realistic description of differential kinematic distributions. In this section, we investigate some important distributions as well as the effects of imposing experimental cuts. The parton-shower simulations are performed using {\sc Pythia6} \cite{Sjostrand:2006za}.

For each benchmark scenario we choose four different parameter points, $(m_{\textrm{DM}}, M_{\textrm{med}})=(1, 10)$, $(1, 500)$, $(500, 10)$ and $(500, 500)$,
all in units of GeV, to present our results. We

\newpage
\begin{figure}[!h]
\centering
\includegraphics[width=0.46\linewidth]{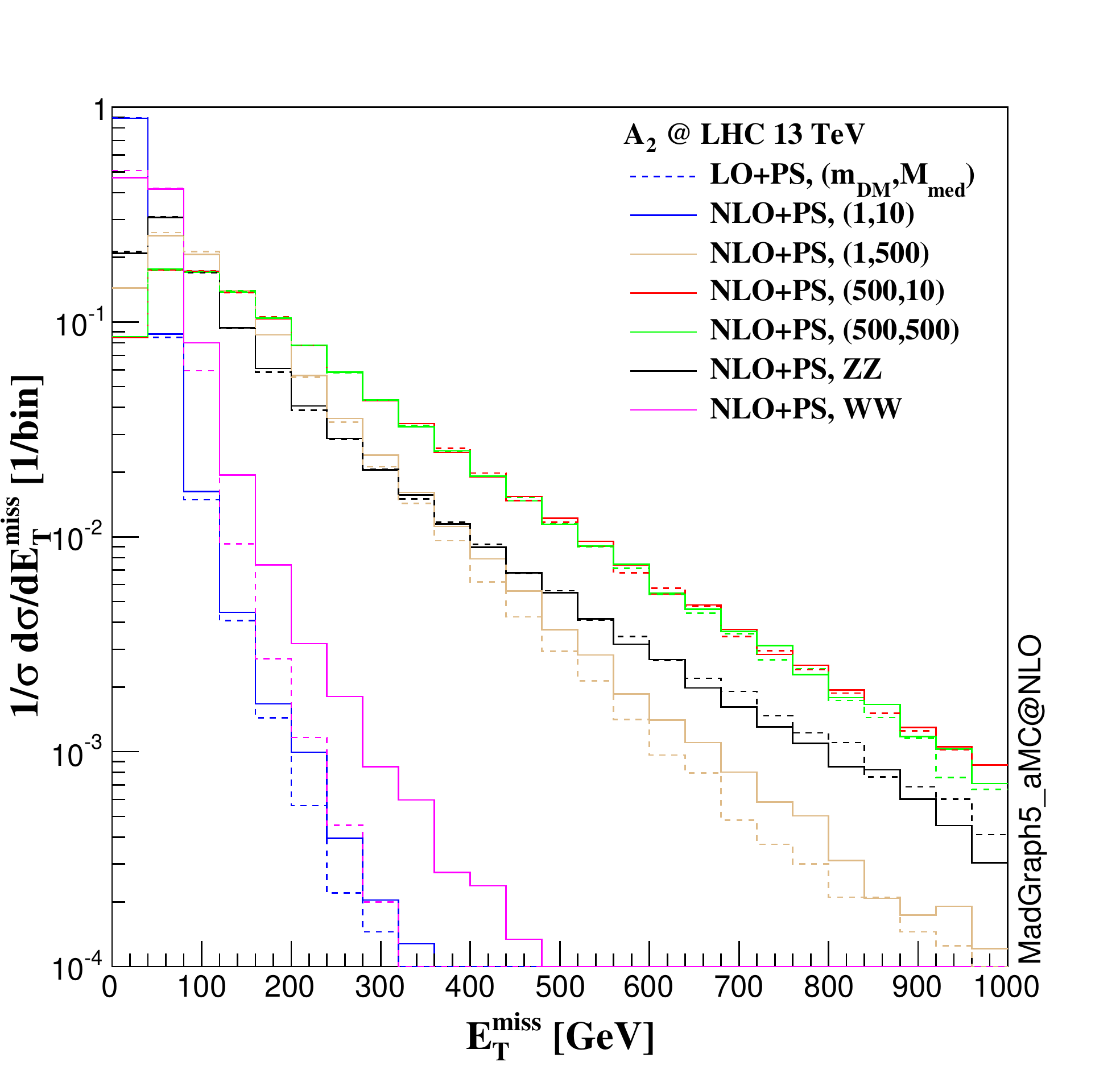}
 \includegraphics[width=0.46\linewidth]{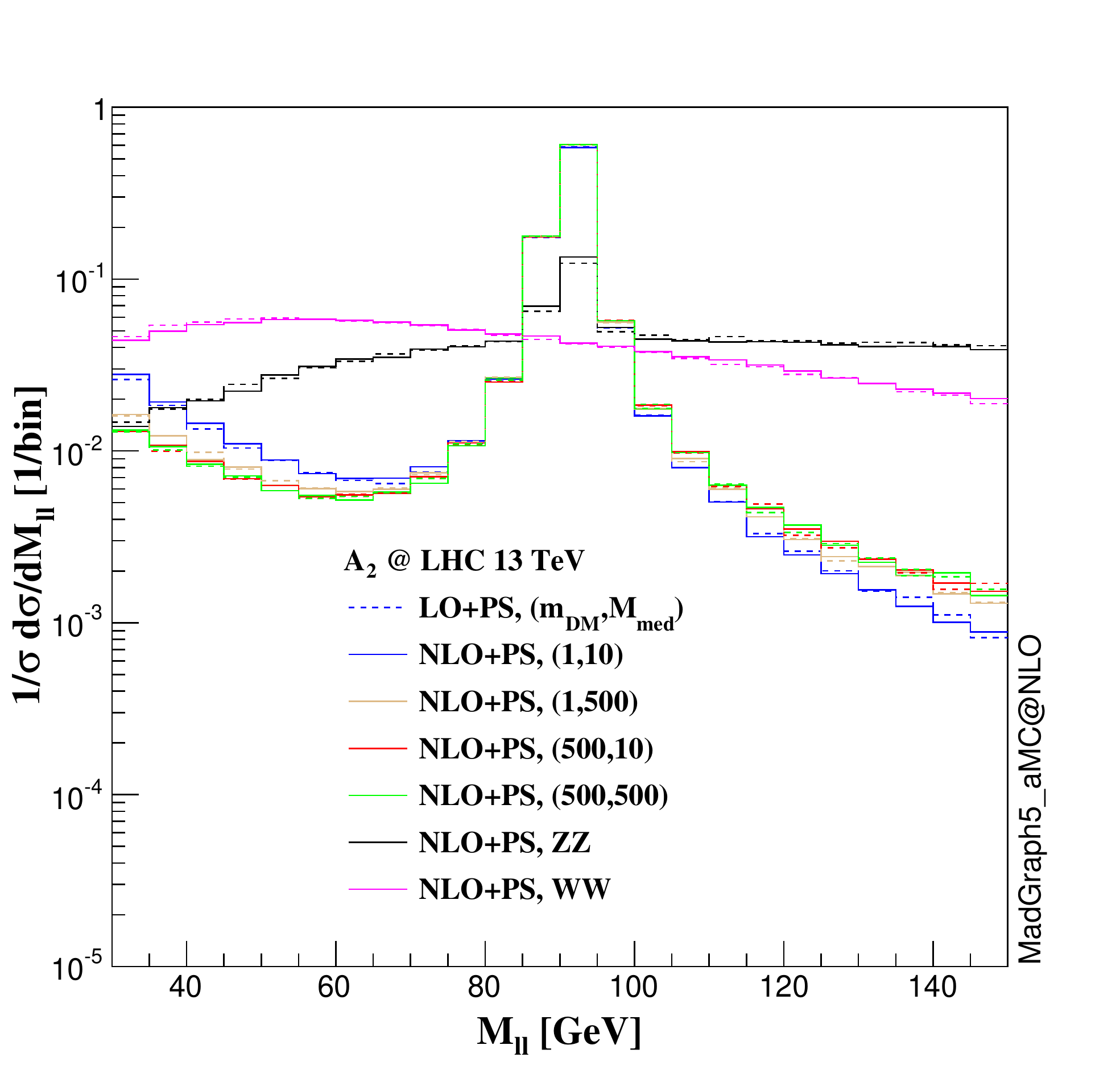} \\
\includegraphics[width=0.46\linewidth]{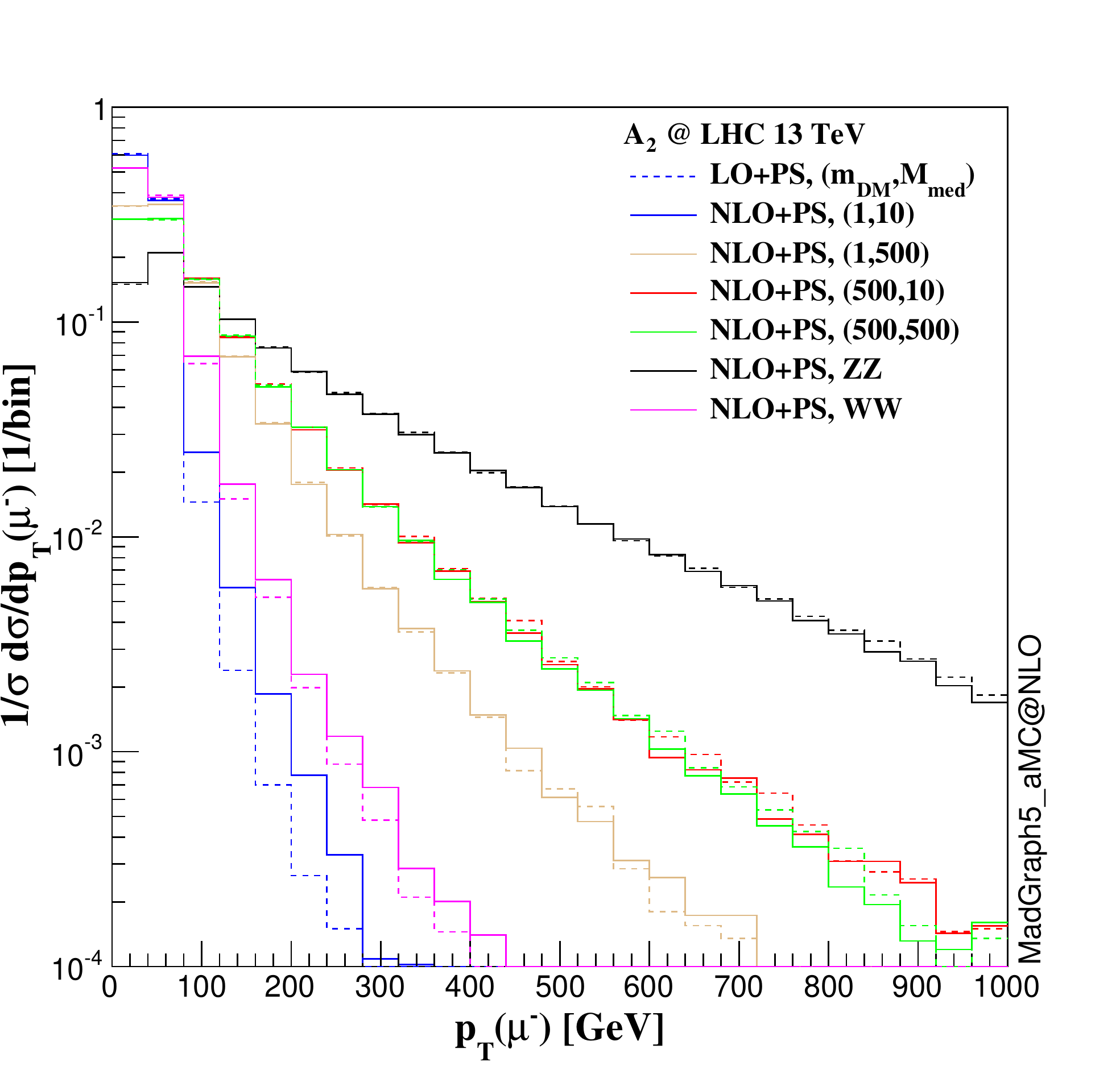}
 \includegraphics[width=0.46\linewidth]{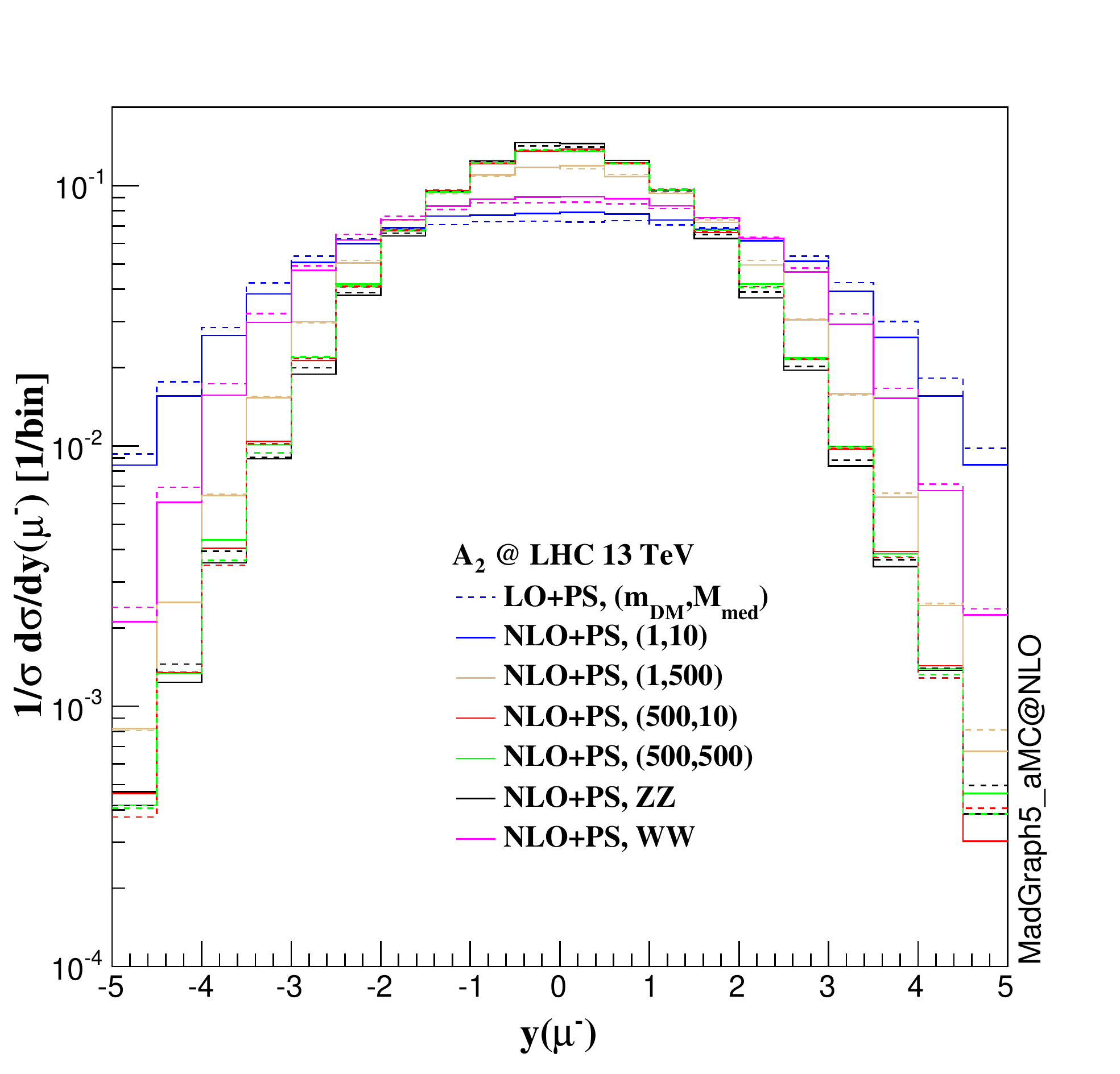} \\
\includegraphics[width=0.46\linewidth]{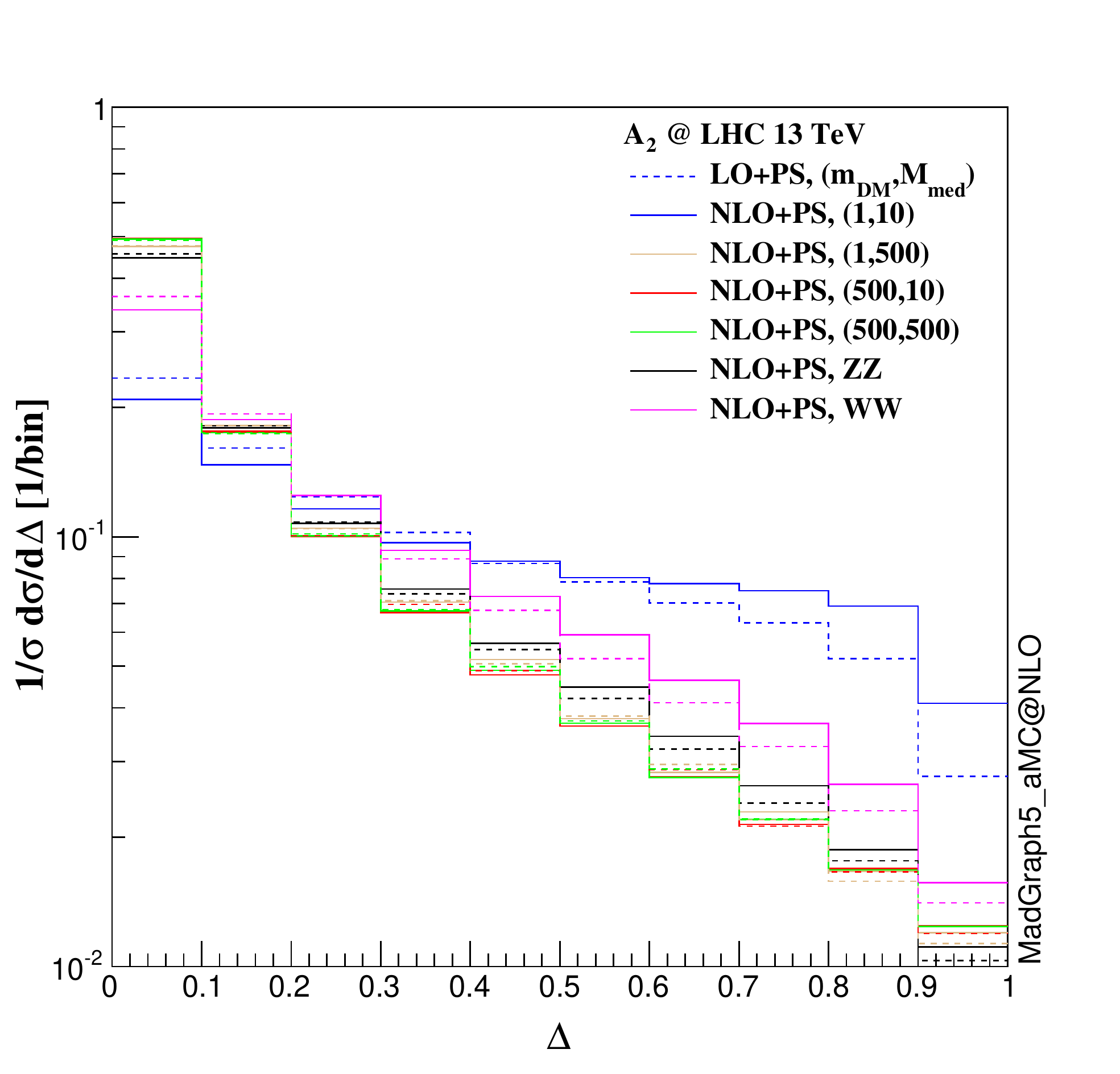}
 \includegraphics[width=0.46\linewidth]{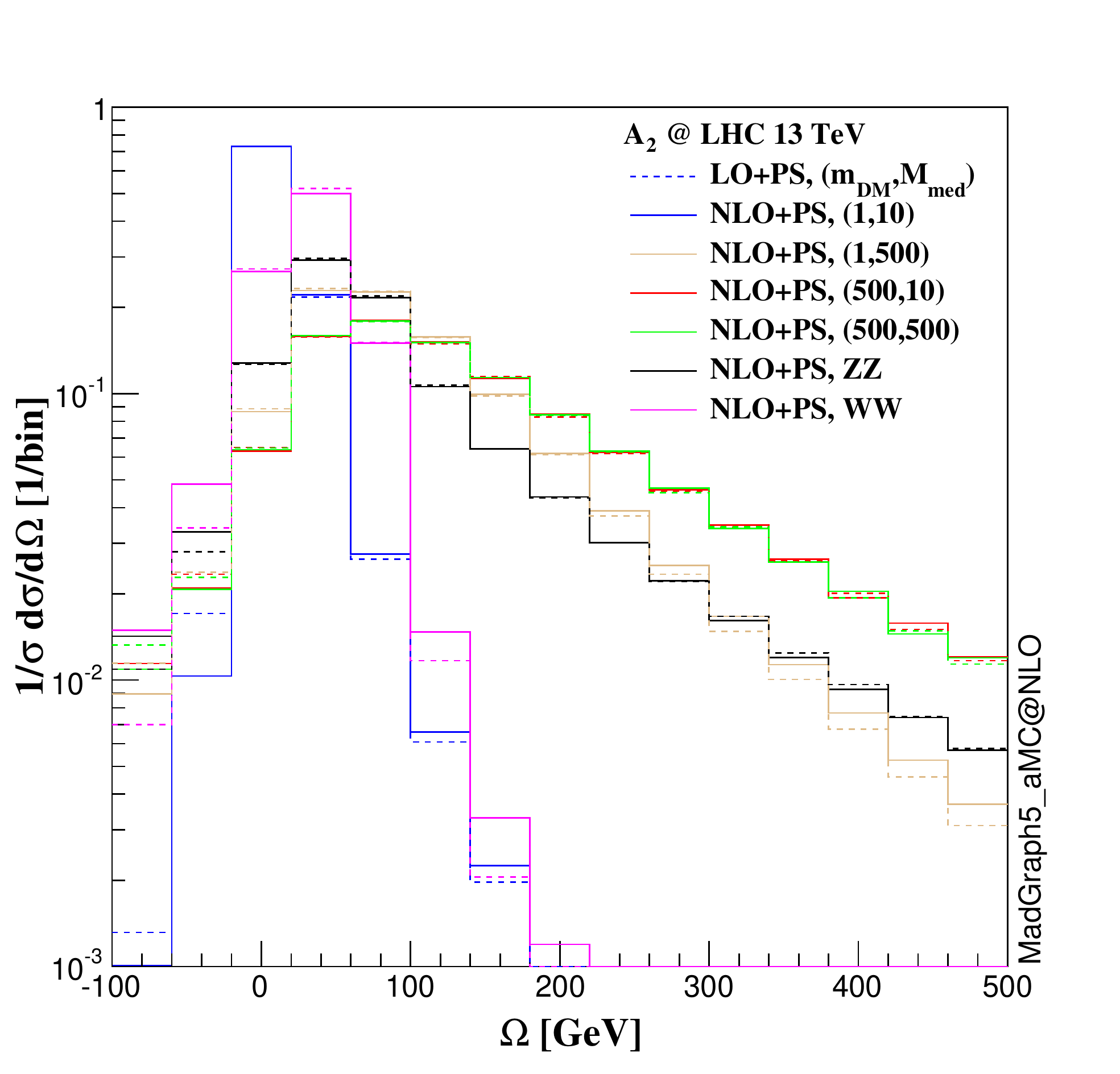} \\
\caption{Kinematic distributions in the benchmark scenario $A_2$ for four different parameter points. The SM background processes $pp\to ZZ, WW\to\nu\bar\nu\mu^+\mu^-$ are also shown. $p_T(\mu^+)$ and $y(\mu^+)$ have similar distributions as $p_T(\mu^-)$ and $y(\mu^-)$, so we do not show them here.}
\label{fig:kin_a2}
\end{figure}

\newpage
\begin{figure}[!h]
\centering
\includegraphics[width=0.46\linewidth]{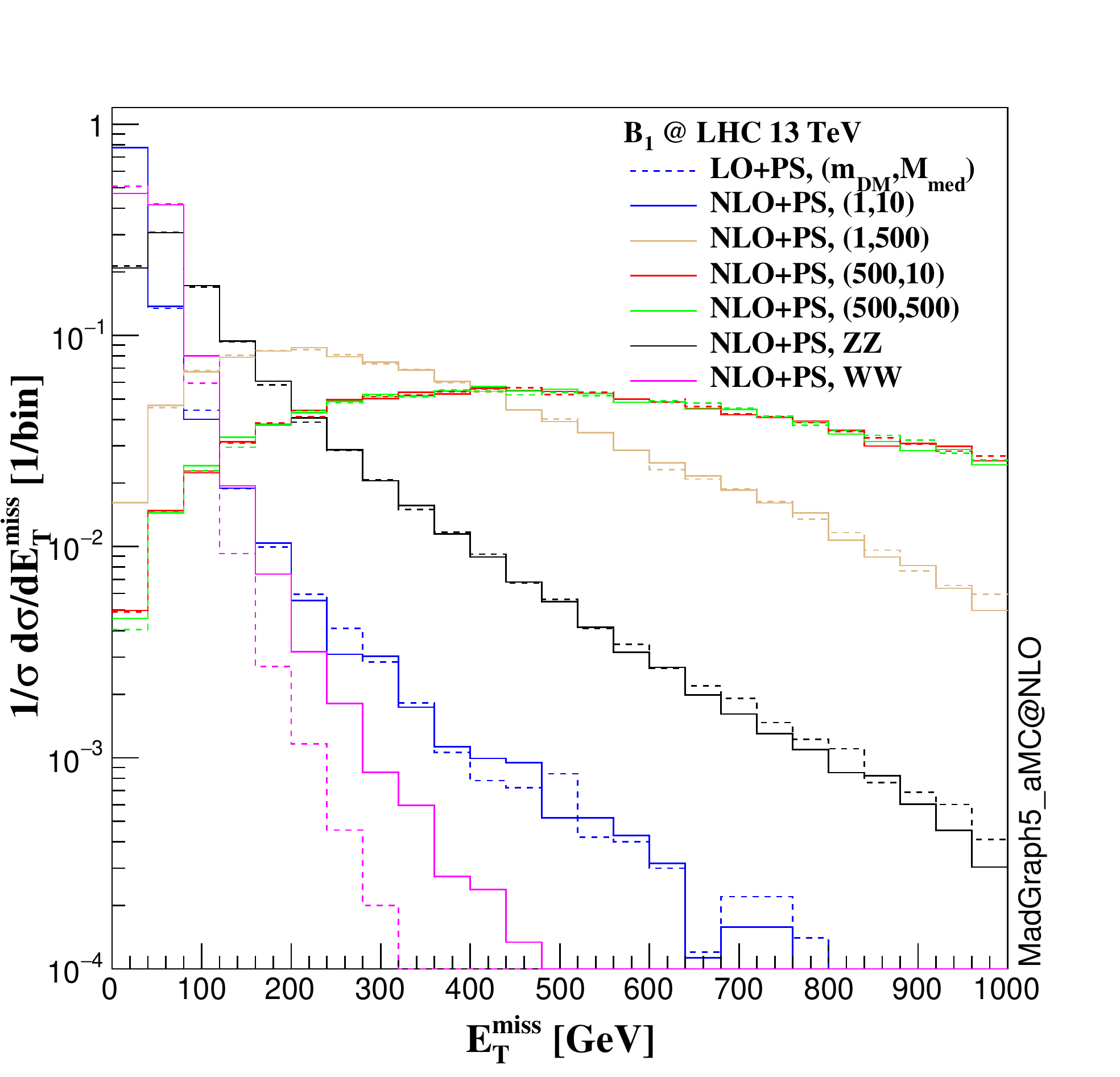}
 \includegraphics[width=0.46\linewidth]{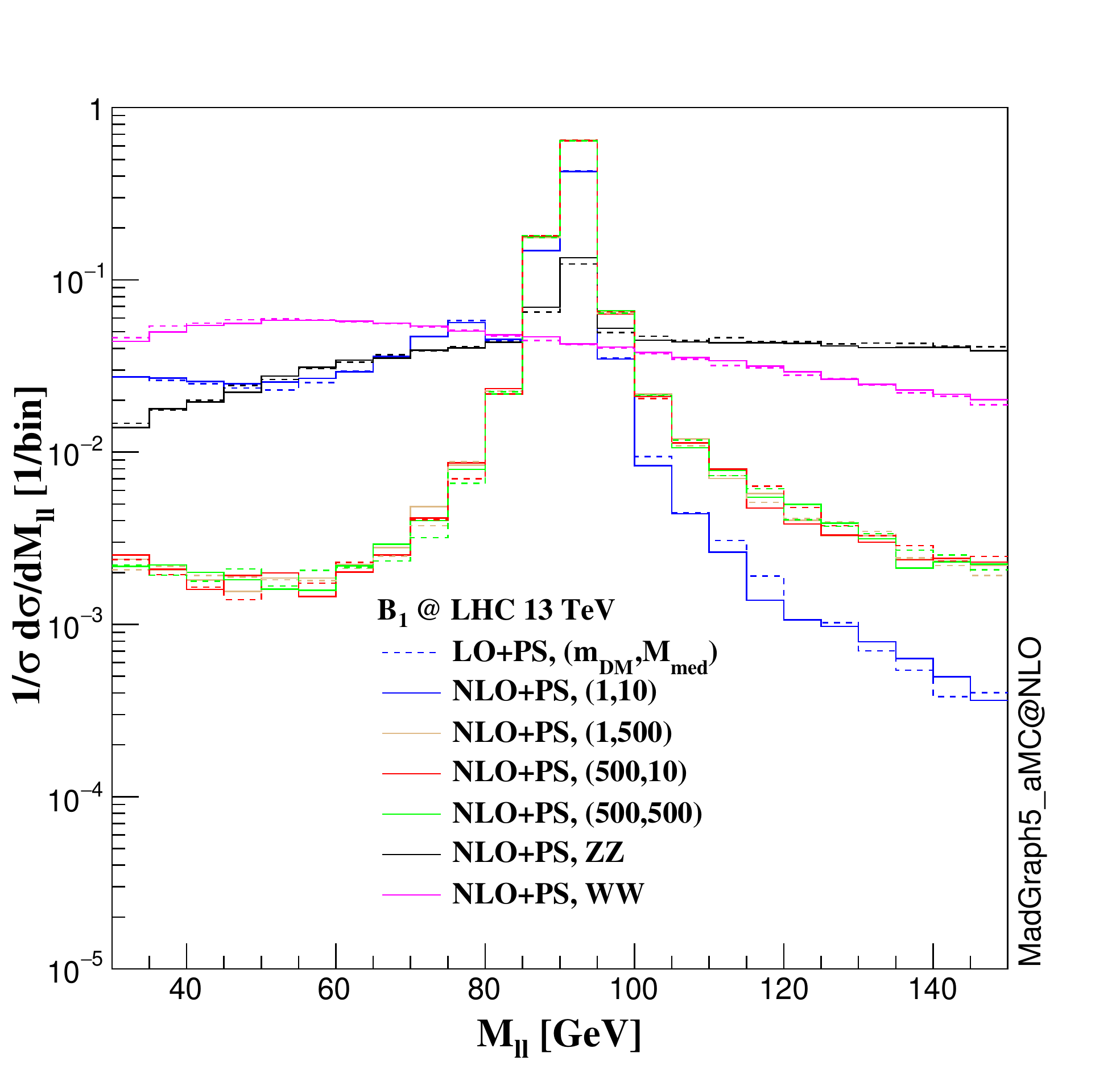} \\
\includegraphics[width=0.46\linewidth]{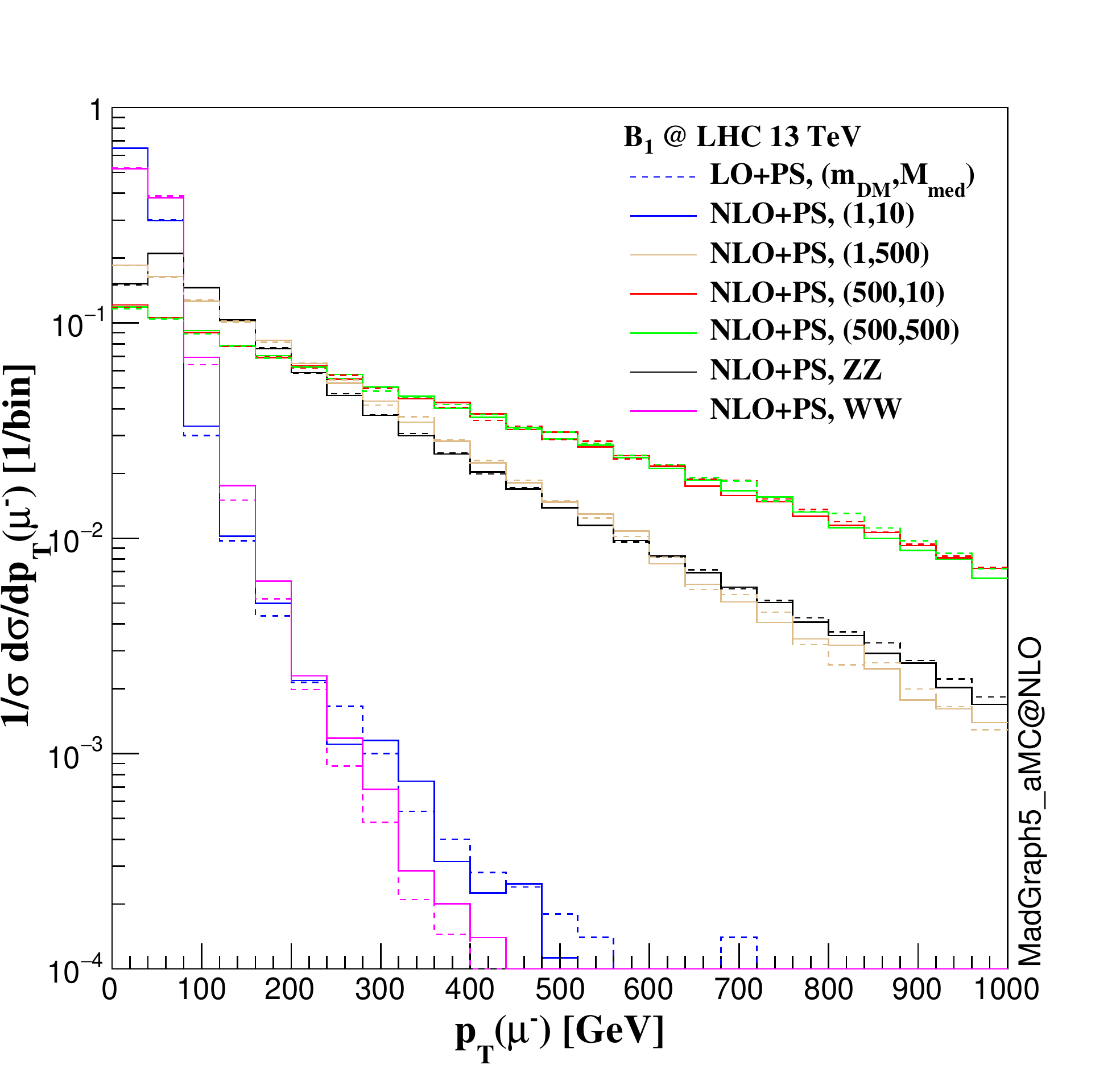}
 \includegraphics[width=0.46\linewidth]{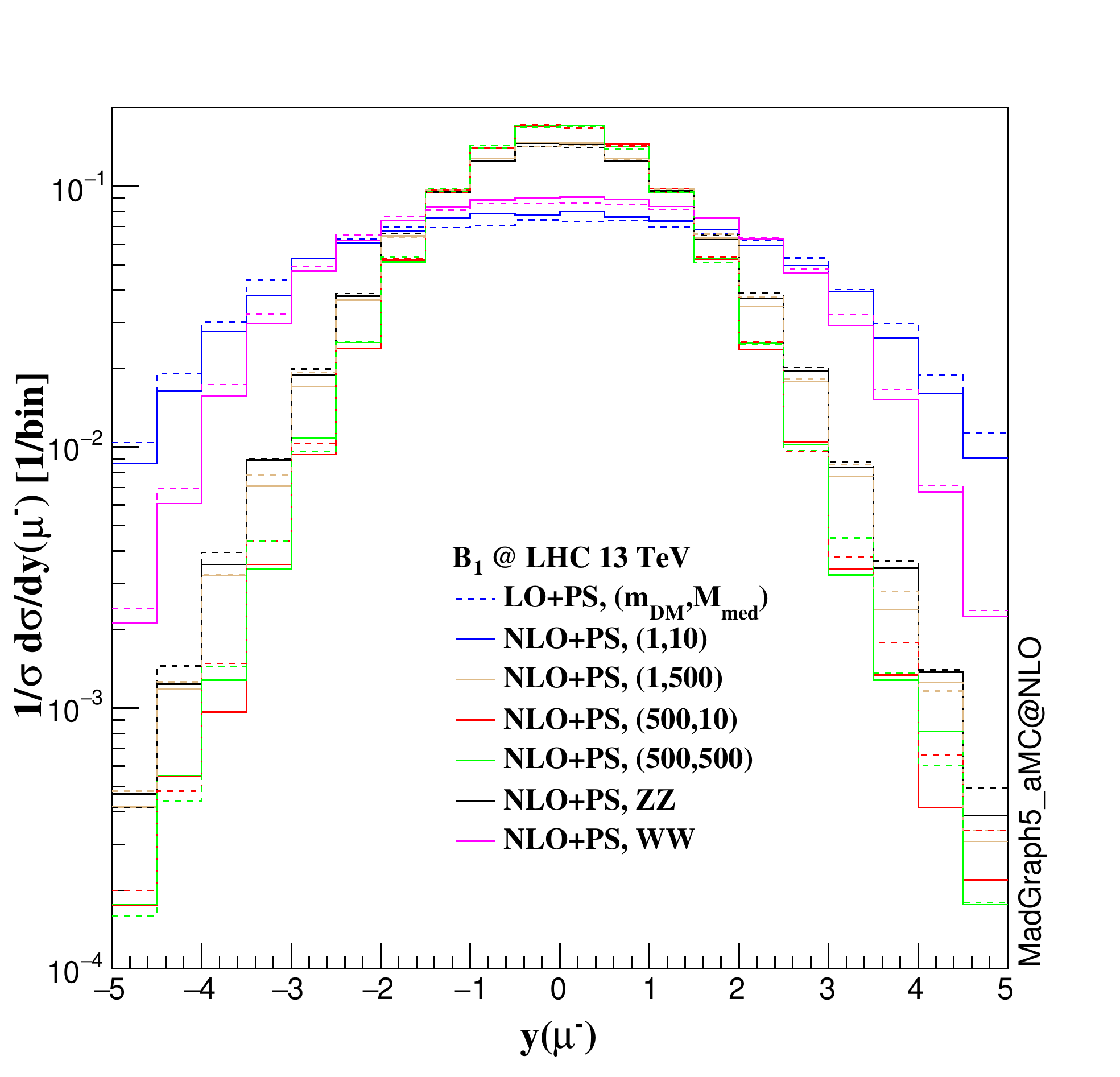} \\
\includegraphics[width=0.46\linewidth]{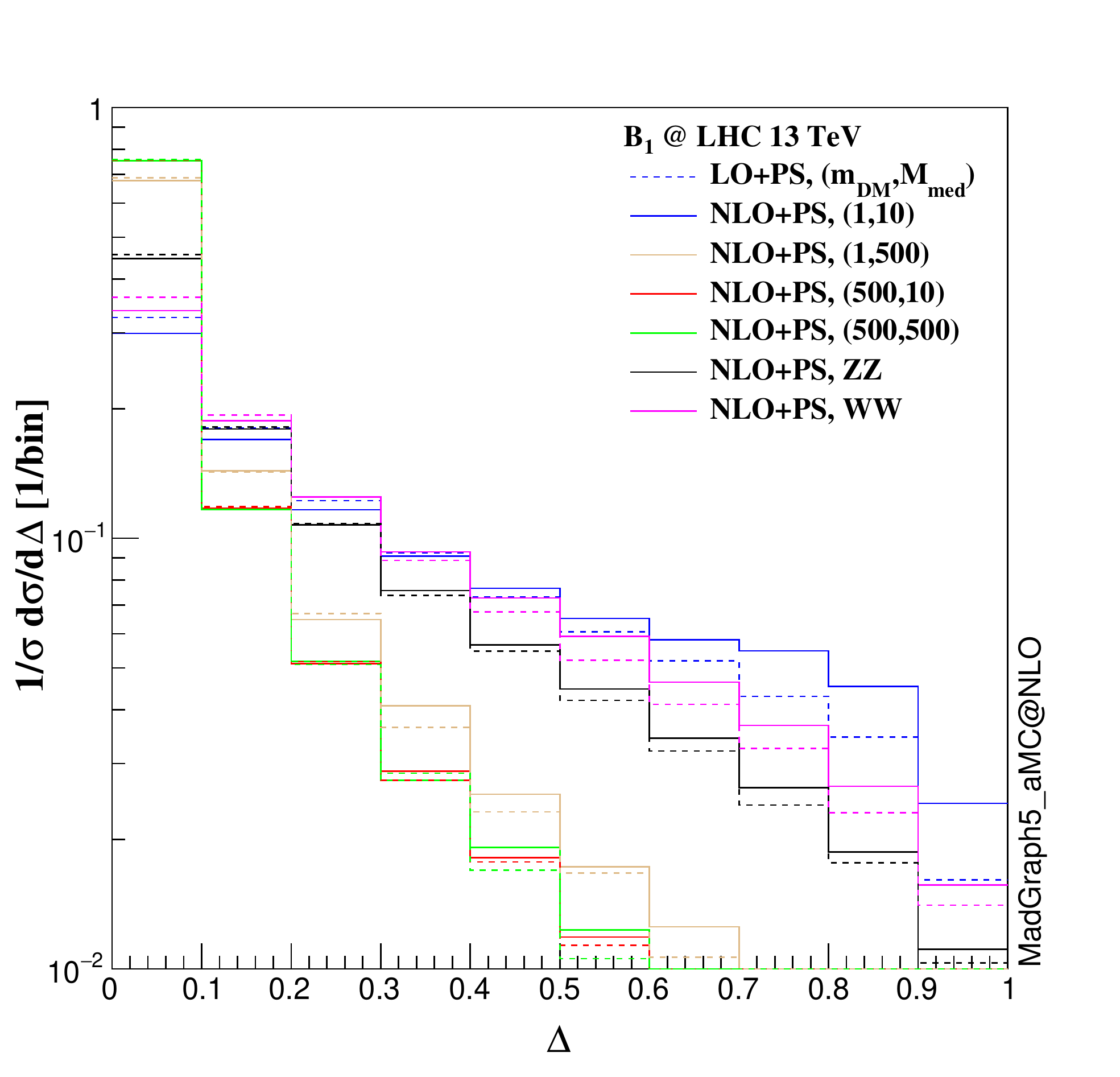}
 \includegraphics[width=0.46\linewidth]{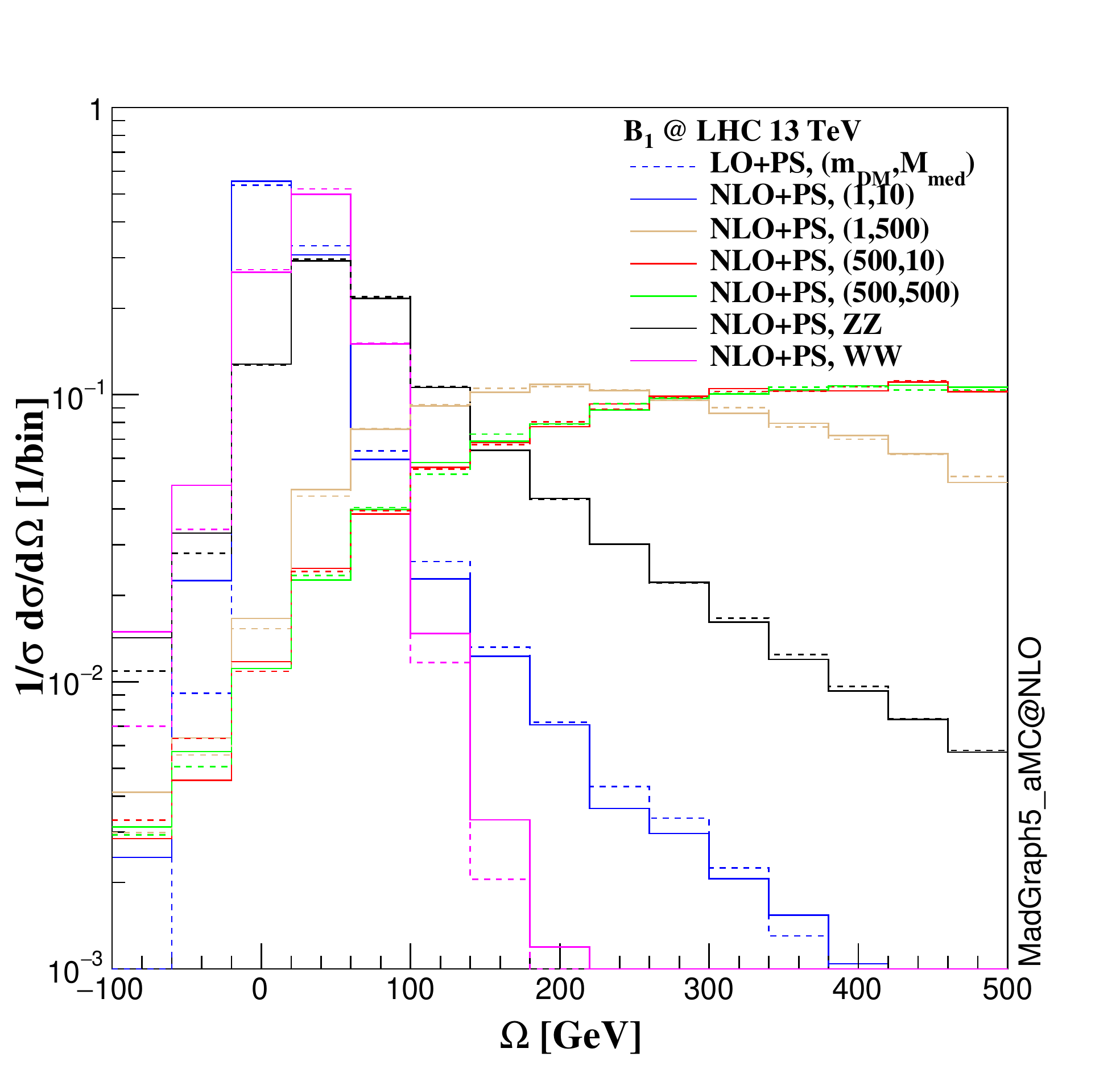} \\
\caption{Kinematic distributions in the benchmark scenario $B_1$ for four different parameter points. The SM background processes $pp\to ZZ, WW\to\nu\bar\nu\mu^+\mu^-$ are also shown. $p_T(\mu^+)$ and $y(\mu^+)$ have similar distributions as $p_T(\mu^-)$ and $y(\mu^-)$, so we do not show them here.}
\label{fig:kin_b1}
\end{figure}

\noindent
study the differential cross sections in six observables: the missing transverse energy $E_T^{\miss}$,
the dilepton invariant mass $M_{ll}$, the muon transverse momentum $p_T(\mu^-)$ and rapidity $y(\mu^-)$, and
the variables $\Delta\equiv |E_T^{\miss}-p_T^{ll}|/p_T^{ll}$ and $\Omega\equiv -\vec{\cancel{p}}_T\cdot\vec{p}_T^{\,ll}/p_T^{ll}$. The observable $\Delta$ measures the difference between the missing transverse energy and the transverse momentum of~the
$Z$-boson candidate, which should be small for the associated production of DM with a $Z$ boson. The observable $\Omega$ depends on the missing transverse momentum and its angle with the direction of the $Z$-boson candidate. Cuts on these variables are very efficient to suppress the backgrounds. For comparison, we show the SM backgrounds $pp\to ZZ\to\nu\bar{\nu}\mu^+\mu^-$ and $pp\to W^+W^-\to\nu\bar{\nu}\mu^+\mu^-$, which also give rise to a mono-$Z$ signal.

Our results for the two benchmark cases $A_2$ and $B_1$ are shown in figures~\ref{fig:kin_a2} and \ref{fig:kin_b1}, respectively. We observe that the $WW$-induced SM background is concentrated in regions of small $E_T^{\miss}$, $p_T(\mu^-)$ and $\Omega$ and has no resonance peaks in the dilepton invariant mass, and hence it should be easy to select proper cuts to suppress this background.
On the other hand, the $ZZ$-induced SM background exhibits similar distributions as the DM signal processes. From the discussion in previous section, we know that the signals with light DM masses and heavy mediators have large cross sections in general because the mediator can be on-shell. To search for this kind of signals, stringent cuts on $E_T^{\miss}$, $M_{ll}$ and $\Omega$ should be imposed,
while other cuts can be kept loose in order to increase the ratio of signal over backgrounds. Notice that the signal in the benchmark scenario $B_1$ appears to have larger $E_T^{\miss}$, $p_T(\mu^-)$ and $\Omega$ than that in the benchmark scenario $A_2$. This is because the benchmark scenario $B_1$ essentially corresponds to the $s$-channel production of a $Z$ boson in association with the mediator $Y_0$, while the benchmark scenario $A_2$ essentially corresponds to the $t$-channel production of a $Z$ boson in association with the mediator $Y_1$. Because of the denominator in the $t$-channel propagator, events containing particles of high transverse momenta are strongly suppressed. In contrast, the $s$-channel process has no such suppression effect, and thus can have relative larger cross sections in regions of high transverse momenta. These three observables can be used as discriminators in case any signal is observed.

\section{\boldmath Discovery potential of the mono-$Z$ signal at the  13 TeV LHC}
\label{sec:lhc}

\begin{table}\centering
\begin{tabular}{c|c|c|c|c|c|c}
\hline\hline
$M_{\textrm{med}}\,[\gev]$ & Basic cuts & $E_T^{\miss}$ & $M_{ll}$ & $\Delta$ & $\Omega$
 & $\epsilon_{\textrm{cut}}$ \\
\hline
\multicolumn{7}{c}{Benchmark scenario $A_2$} \\
\hline
100 & 326 & 275 & 158 & 53.7 & 20.4 & 0.061 \\
200 & 97.7 & 86.6 & 59.1 & 28.0 & 11.7 & 0.117 \\
500 & 12.9 & 12.0 & 9.39 & 5.76 & 2.73 & 0.209 \\
1000 & 1.68 & 1.59 & 1.32 & 0.890 & 0.451 & 0.265 \\
\hline
$ZZ$ & 4747 & 3688 & 2101 & 1379 & 16.0 & $3.24\cdot 10^{-3}$ \\
$WW$ & 988 & 479 & 82.6 & 10.6 & 0.487 & $4.31\cdot 10^{-4}$ \\
\hline
\multicolumn{7}{c}{Benchmark scenario $B_1$} \\
\hline
100 & 0.966 & 0.897 & 0.684 & 0.466 & 0.238 & 0.245 \\
200 & 0.331 & 0.319 & 0.281 & 0.228 & 0.129 & 0.388 \\
500 & $5.09\cdot 10^{-2}$ & $5.02\cdot 10^{-2}$ & $4.81\cdot 10^{-2}$ & $4.35\cdot 10^{-2}$
 & $2.78\cdot 10^{-2}$ & 0.546 \\
1000 & $6.81\cdot 10^{-3}$ & $6.77\cdot 10^{-3}$ & $6.63\cdot 10^{-3}$ & $6.23\cdot 10^{-3}$
 & $4.24\cdot 10^{-3}$ & 0.622 \\
\hline
$ZZ$ & 4747 & 3688 & 2101 & 1379 & 16.0 & $3.24\cdot 10^{-3}$ \\
$WW$ & 988 & 479 & 82.6 & 10.6 & 0.487 & $4.31\cdot 10^{-4}$ \\
\hline\hline
\end{tabular}
\caption{Cross sections (in $\fb$) for mono-$Z$ production in the channel $pp\to Z(\to\mu^+\mu^-)+\cancel{E}_T$ at the 13 TeV LHC after a series of kinematic cuts, for the benchmark scenarios $A_2$ (top) and $B_1$ (bottom). The last column shows the cut acceptance $\epsilon_{\textrm{cut}}$. We assume $m_{\textrm{DM}}=10~\gev$.}
\label{tab:cut1}
\end{table}

In this section we study the discovery potential of the mono-$Z$ signal in the $Z\to\mu^+\mu^-$ channel at the 13 TeV LHC. In general, the production cross section is a function of the relevant couplings, the DM mass and the mediator mass. For illustration purposes, we fix the coupling constants according to the suggestions of the ATLAS/CMS DM forum and take $m_{\textrm{DM}}=10~\gev$, as for light DM the LHC experiments have better sensitivities than direct-detection experiments. The mediator mass is varied between 100~GeV and 1000~GeV. Based on the kinematic distributions studied in the previous section, we first impose the basic cuts
\begin{equation}
    p_T(\mu^{\pm}) > 20~\gev \,, \qquad |y(\mu^{\pm})| < 2.5
\end{equation}
to select the signal events and reduce the SM background. We then apply, step after step, a series of more advanced cuts, namely
\begin{equation}
   E_T^{\miss} > 100~\gev \,, \quad M_{ll} \in [85~\gev, 100~\gev] \,, \quad
   \Delta < 0.4 \,, \quad \Omega > 80~\gev \,.
\end{equation}

The effective cross sections of the signal and SM backgrounds obtained after applying these various cuts are given in table~\ref{tab:cut1} for the benchmark scenarios $A_2$ and $B_1$. We can see that the backgrounds are efficiently suppressed after the advanced cuts applied. From the cut acceptances of the signals, we find that the events with heavier  mediator mass can pass the selection cuts more easily. More specifically, the cut acceptances in the benchmark scenario $B_1$ are higher than those in the  benchmark scenario $A_2$.

\begin{figure}\centering
\includegraphics[width=0.5\linewidth]{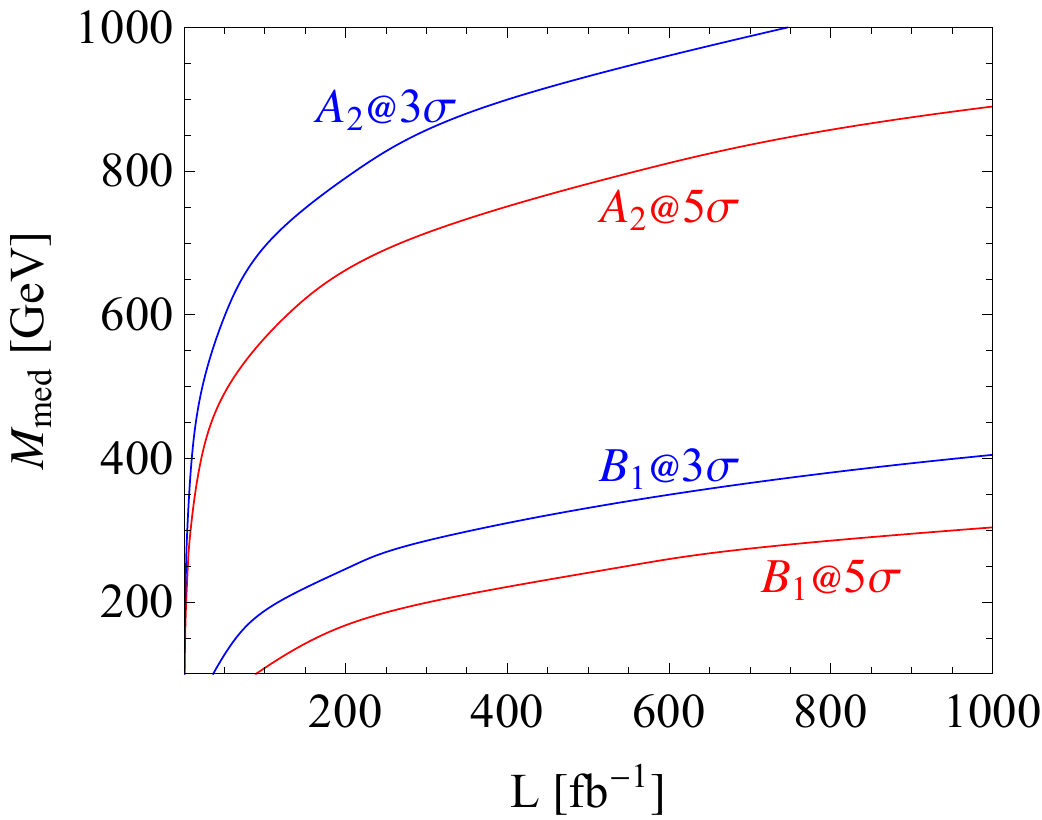}
\caption{$5\sigma$ discovery and $3\sigma$ exclusion limits for DM production at the LHC. If a discovery is made, then the regions below the red lines are favored. If no signal is found, then the regions below the blue lines are excluded. Results in the benchmark scenario $B_1$ are given for $\Lambda=1$~TeV.}
\label{fig:lhc1}
\end{figure}

Given the cross sections obtained after applying all cuts, we can estimate the discovery or exclusion potentials that can be expected in Run-II of the LHC. The corresponding results are shown in figure~\ref{fig:lhc1}. If no signal events will have been detected after accumulating an integrated luminosity of $750~\fb^{-1}$, then the regions with mediator mass $M_{\textrm{med}}<1000~\gev$ can be excluded at the $3\sigma$ level, assuming $m_{\textrm{DM}}=10$~GeV in the benchmark scenario $A_2$. On the other hand, if the mediator mass satisfies $M_{\textrm{med}}<890~\gev$, then the mono-$Z$ signal is going to be discovered at the $5\sigma$ level before accumulating an integrated luminosity of $1000~\fb^{-1}$. The discovery potential of the mono-$Z$ signal in the benchmark scenario $B_1$ is also shown in figure~\ref{fig:lhc1}. Because of the small cross section, the discovery potential is not as promising as that for scenario $A_2$. In order to show the two cases in one plot, we have multiplied the cross section in benchmark scenario $B_1$ by a factor of 9, corresponding to lowering the new-physics scale $\Lambda$ from 3~TeV to 1~TeV
\footnote{In the case of $M_{\textrm{med}}<500~\gev$, the change of  $\Lambda$ from 3~TeV to 1~TeV has negligible effect on the total decay width of $Y_0$
since the dominant decay channel is $Y_0 \to X_D\overline{X}_D$.}.
In this case, the LHC with an integrated luminosity of $1000~\fb^{-1}$ can probe or exclude the mediator with a mass lighter than about 300~GeV or 400~GeV, respectively, again assuming $m_{\textrm{DM}}=10$~GeV.

\section{Conclusions and outlook}
\label{sec:Conclusion}

In this work, we have implemented a class of simplified models for DM production via $s$-channel vector or scalar mediators in the {\sc FeynRules/MadGraph5\_aMC@NLO} framework, which allows us to obtain accurate and realistic predictions for DM production rates as well as kinematic distributions for Run-II of the LHC. Apart from interactions already presented in \cite{Backovic:2015soa}, we have added and validated direct interactions between the mediators and the electroweak sector of the SM. Our implementation provides a theoretical basis for future DM searches at the LHC.

As an illustration, we have presented the first NLO QCD predictions for mono-$Z$ signals in simplified models including parton-shower effects. We have considered several benchmark scenarios including both spin-0 and spin-1 mediators. Our calculation has been fully automated and can readily be reproduced. We have presented predictions for the total $pp\to Z+\cancel{E}_T$ production cross sections for the parameter sets suggested by the ATLAS/CMS DM forum in \cite{Abercrombie:2015wmb}. The $K$-factors vary in the range of about $1.3-1.5$, with the precise values depending on the DM and mediator masses. This shows that the NLO corrections have a noticeable impact on the mono-Z signal and should not be ignored. The theoretical predictions of the cross sections become more reliable at NLO and in many cases the scale uncertainties are reduced.

In a next step we have studied various kinematic distributions in order to better understand the feature of the mono-$Z$ signal. Different coupling structures, DM masses, and mediator masses can all affect these distributions. From these investigations we have obtained important information on suitable cuts needed to select the signal events and suppress backgrounds in the search for the mono-$Z$ signal. If a DM signal is observed in future, some of these variables may be used to determine the structure of the DM couplings. We have also estimated the discovery potential of the mono-$Z$ signal at the 13 TeV LHC in several benchmark scenarios.

The present work can be extended in several ways. First, one should investigate more signal channels in the context of simplified models, such as mono-$W$, mono-$\gamma$ and mono-$h$. Different channels are sensitive to different operators in the simplified models. In a combined study one would thus be able to obtain comprehensive understanding of the DM interactions. Second, the comparison with present experimental results at the 8 TeV LHC, including the mono-$X$ searches as well as dijet resonance searches, would help to constrain the parameter space. Third, from our studies we find that the kinematic distributions in different scenarios can be rather different. Thus, optimized cuts should be chosen in searching for the signals or imposing constraints for specific operators. These interesting problems are left to future work.

\acknowledgments
We would like to thank Mihailo Backovic, Michael Kr\"amer, Fabio Maltoni, Antony Martini, Kentarou Mawatari and Mathieu Pellen for useful discussions. This work was supported by the Cluster of Excellence {\it Precision Physics, Fundamental Interactions and Structure of Matter} (PRISMA-EXC 1098). M.N.~is also supported by the Advanced Grant EFT4LHC of the European Research Council (ERC),
and grant 05H12UME of the German Federal Ministry for Education and Research. The work of C.Z.~was supported by U.S.~Department of Energy under Grant No.~DE-AC02-98CH10886.

\newpage
\appendix
\section{Contact interactions in EFT}
\label{app:eft}

\begin{figure}[h]\centering
\includegraphics[width=0.37\linewidth]{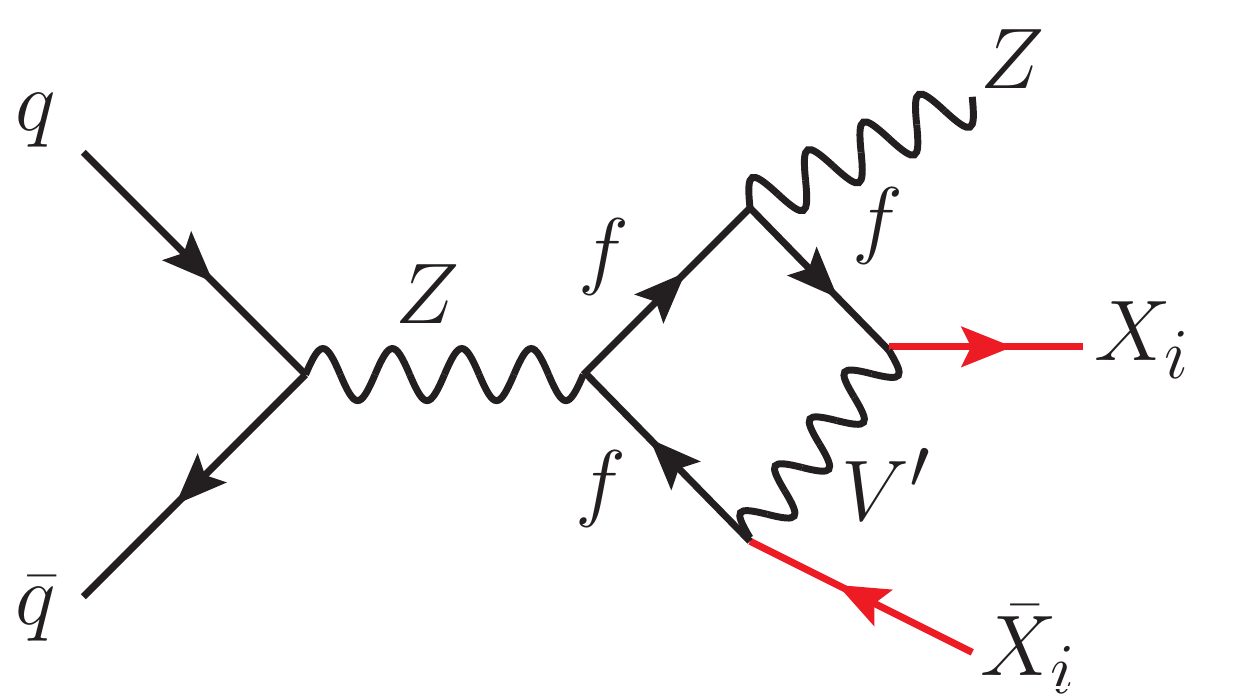}
\caption{Mono-$Z$ production through a loop-induced operator. $f$ is some very heavy fermion, while $V'$ is a very heavy boson.}
\label{fig:monozbox}
\end{figure}

In the main body of this paper we have focused on simplified models for DM interactions with the SM. However, as mentioned in section~\ref{sec:model}, there is also a case in which the simplified-model framework does not apply and an EFT approach should be used. In this appendix we provide some results for the particular case where effective local operators describe contact interactions between DM particles and the electroweak sector of the SM without mediators. Figure~\ref{fig:monozbox} shows an example of a loop diagram containing a new, very heavy particles $f$ and $V'$, which can generate such operators. They could produce signals with electroweak gauge bosons and Higgs bosons in the final states \cite{Carpenter:2012rg,Alves:2015dya,Crivellin:2015wva}, and therefore should be  investigated as well \cite{Abercrombie:2015wmb}.

The complete set of operators up to dimension seven can be found in \cite{Cotta:2012nj,Carpenter:2012rg,Crivellin:2015wva}. A comprehensive study including the full set of operators is beyond the scope of this work. Instead, we only focus on two of them. Our main purpose is to present sample results for such loop-induced operators and demonstrate that the approach we have used for simplified models can easily be adapted to the case of EFT studies. In particular, we consider the operators
\begin{align}\label{eq:eft}
   {\cal L}_{{\textrm{EFT}}}
   = \frac{c^S_{h}}{\Lambda^3} \left(D^\mu\phi^\dagger D_\mu\phi\right) \bar X_D X_D
	+ \frac{c^S_{W}}{\Lambda^3}\,W_{\mu\nu}W^{\mu\nu} \bar X_D X_D \,.
\end{align}
After electroweak symmetry breaking, these operators contain the dimension-5 operator $\bar\chi\chi W^{+\mu}W^-_\mu$ and the dimension-7 operator $\bar\chi\chi W_{\mu\nu}^i W^{i,\mu\nu}$, both of which are discussed as benchmark models in \cite{Abercrombie:2015wmb}. In fact, the interactions in (\ref{eq:eft}) are equivalent to those in the benchmark scenarios $B_1$ and $B_3$, apart from not having a mediator. Other operators, including the dipole operator $\bar X_D\sigma_{\mu\nu} B^{\mu\nu} X_D$, will not be considered here.

\begin{table}
\small\centering
\begin{tabular}{c|cccccccc}
\hline\hline
$m_{\textrm{DM}}$\,[GeV] & 1 & 10 & 50 & 100 & 200 & 400 & 800 & 1300 \\
\hline
\multicolumn{9}{c}{Benchmark scenario $C_1$} \\
\hline
$\sigma(\mathrm{NLO})$\,[pb] & 1.3e-11 &  1.2e-11 & 8.7e-12 & 5.8e-12 & 2.9e-12 & 8.4e-13
 & 9.9e-14 & 7.8e-15 \\
\hline
$K$ & 1.32 & 1.33 & 1.31 & 1.30 & 1.30 & 1.24 & 1.10 & 0.97 \\
\hline
\multicolumn{9}{c}{Benchmark scenario $C_2$} \\
\hline
$\sigma(\mathrm{NLO})$\,[pb] & 6.2e-8 &  6.2e-8 & 6.0e-8 & 5.5e-8 & 4.3e-8 & 2.3e-8
 & 5.4e-9 & 7.3e-10 \\
\hline
 $K$ & 1.22 & 1.23 & 1.24 & 1.22  & 1.21 & 1.19 & 1.06 & 0.91 \\
\hline\hline
\end{tabular}
\caption{The NLO total cross sections (in pb) and $K$-factors for mono-$Z$ production in the channel $pp\to Z(\to\mu^+\mu^-)+\cancel{E}_T$ at the 13 TeV LHC in the benchmark scenarios $C_1$ (top) and $C_2$ (bottom). We use the short-hand notation ``e-$n$'' for $10^{-n}$.}
 \label{tab:nlo7}
\end{table}

We have implemented these two operators following the procedure described in section~\ref{sec:mg}. We shall consider two benchmark scenarios $C_1$ and $C_2$, corresponding to ($c_{h}^S=1$, $c_{W}^S=0$) and ($c_{h}^S=0$, $c_{W}^S=1$), respectively. Following the recommendation of \cite{Abercrombie:2015wmb}, we take $\Lambda=3$~TeV and scan over seven DM mass points: $m_\chi=1, 10, 50, 100, 200, 400, 800, 1300$, all in units of GeV. The resulting cross sections scale like $\sigma\sim\Lambda^{-6}$, so it is easy to translate our results to lower values of the new-physics scale $\Lambda$. The total cross sections and $K$-factors are presented in table~\ref{tab:nlo7}. We see that for $\Lambda=3$~TeV the predicted cross sections are extremely small and would be impossible to measure at the LHC. However, although the energy scale $\Lambda\sim 3$~TeV may thus not be explored, it is still possible to probe lower values of $\Lambda$. The difference in the Lorentz structures for scenarios $C_1$ and $C_2$ leads to significant differences in the production cross sections. We also find that the $K$-factors in the benchmark scenario $C_i$ tend to be somewhat smaller than those in benchmark scenarios $A_i$ and $B_i$.

\begin{figure}\centering
\includegraphics[width=0.5\linewidth]{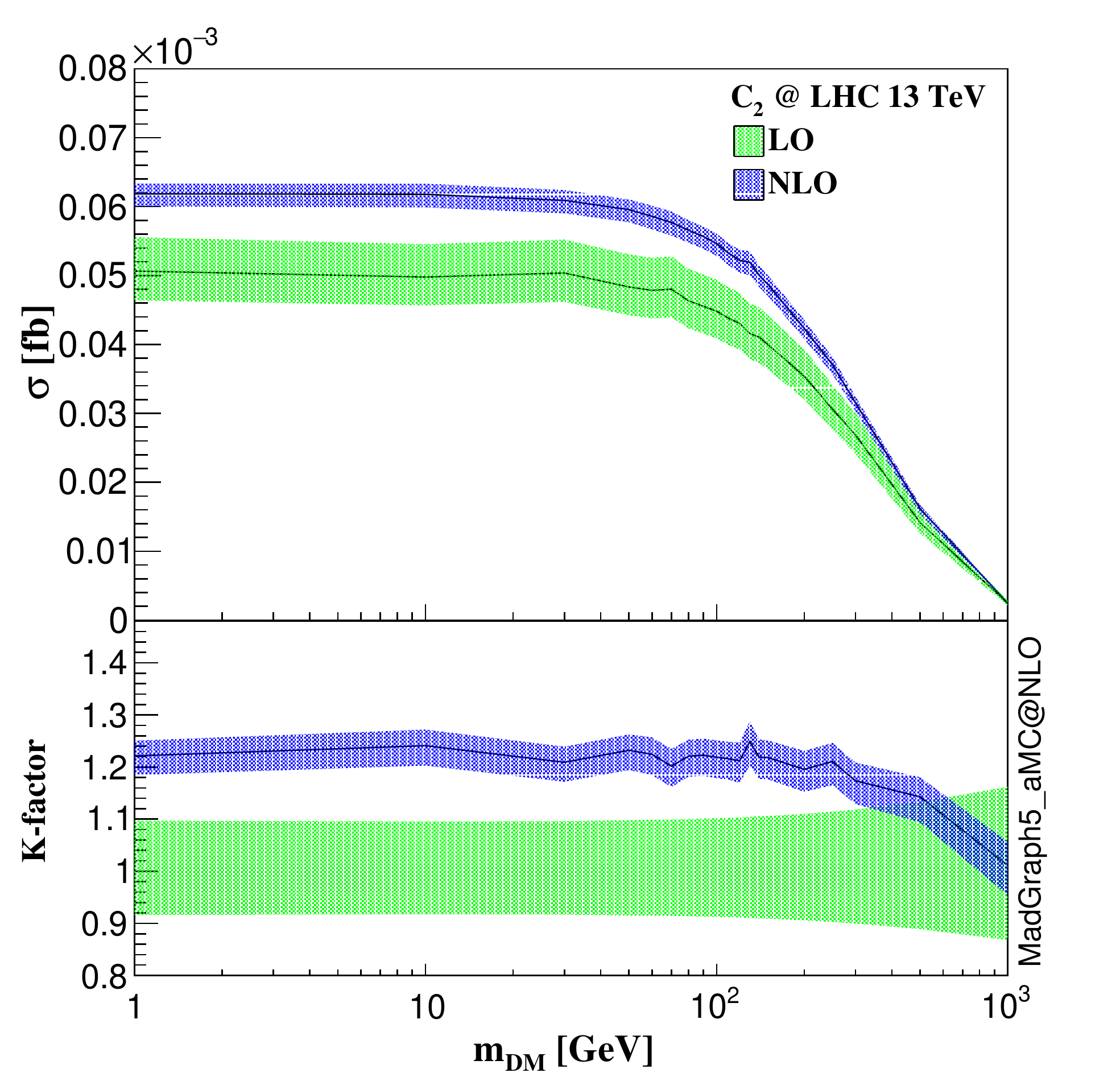}
\caption{Benchmark scenario $C_2$ with $\Lambda=3$~TeV: The $pp\to Z(\to\mu^+\mu^-)+\cancel{E}_T$ cross section and $K$-factor as functions of the DM mass. The  bands represent the scale uncertainties, as estimated by varying the scales $\mu_f$ and $\mu_r$ independently by a factor~2 about their default values.}
\label{fig:dmmed_c2}
\end{figure}

In figure~\ref{fig:dmmed_c2}, we show the cross section in the benchmark scenario $C_2$, which is not quite as suppressed as the cross section in scenario $C_1$, as a function of the DM mass. It is interesting to compare this figure with the left plot of figure~\ref{fig:dmmed_b1}. As explained earlier, the cross section in EFT drops much slower than that in simplified models. Note also that the scale uncertainties are significantly reduced at NLO. Because of the smallness of the signal cross section we refrain from showing detailed kinematic distributions for the scenarios $C_{1,2}$.

\begin{table}
\centering
\begin{tabular}{c|c|c|c|c|c|c}
\hline\hline
 & basic & $E_T^{\miss}$ & $M_{ll}$ & $\Delta$ & $\Omega$ & $\epsilon_{\textrm{cut}}$ \\
\hline
$C_2$& $6.13\cdot 10^{-5}$ & $6.08\cdot 10^{-5}$ & $5.92\cdot 10^{-5}$ & $5.56\cdot 10^{-5}$
 & $3.82\cdot 10^{-5}$ & 0.622 \\
\hline
$ZZ$ & 4747 & 3688 & 2101 & 1379  & 16.0  &  $3.24\cdot 10^{-3}$ \\
$WW$ & 988  & 479  & 82.6 & 10.6  & 0.487 &  $4.31\cdot 10^{-4}$ \\
\hline\hline
\end{tabular}
\caption{Benchmark scenario $C_2$ with $\Lambda=3$~TeV: The cross sections (in $\fb$) for mono-$Z$ production in the channel $pp\to Z(\to\mu^+\mu^-)+\cancel{E}_T$ at the 13 TeV LHC after a series of cuts. The last column shows the cut acceptance $\epsilon_{\textrm{cut}}$. We assume $m_{\textrm{DM}}=10~\gev$.}
\label{tab:cut3}
\end{table}

\begin{figure}
\centering
\includegraphics[width=0.5\linewidth]{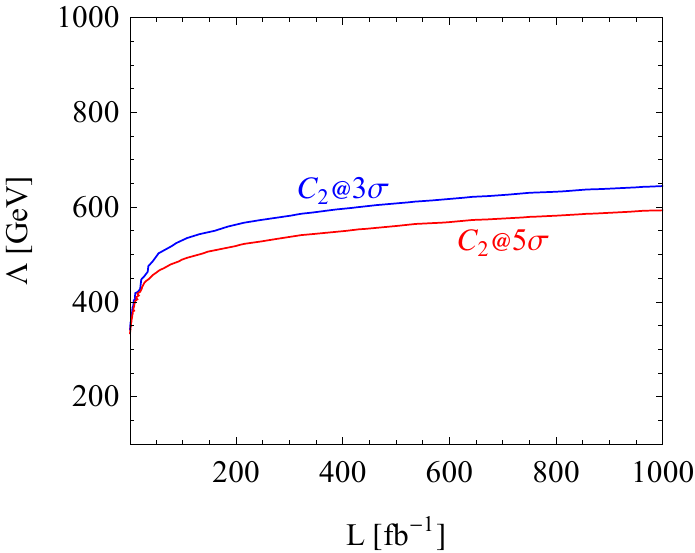}
\caption{$5\sigma$ discovery and $3\sigma$ exclusion limits for DM production at the LHC. If a discovery is made, then the regions below the red lines are favored. If no signal is found, then the regions below the blue lines are excluded.}
\label{fig:lhc3}
\end{figure}

The cross sections of the signal and backgrounds for benchmark scenario $C_2$, obtained after applying the same set of cuts as described in section~\ref{sec:kin}, are given in table~\ref{tab:cut3}. With $\Lambda=3$~TeV the cross sections are orders of magnitude too small for detecting a signal at the LHC. The corresponding discovery potential and exclusion limits are shown in figure~\ref{fig:lhc3}. The energy scale in EFT can be constrained to be larger than 650 GeV if no signal is observed at the 13 TeV LHC (with an integrated luminosity of $1000~\fb^{-1}$). This constraint does not improve significantly even if more data are accumulated. On the other hand, for such a low new-physics scale the application of the EFT framework is highly questionable.

\newpage
\section{Cross sections and $K$-factors in benchmark scenarios A$_{1,3}$ and B$_{2,3}$}
\label{app:results}

In tables~\ref{tab:nlo1}--\ref{tab:nlo6} we present our results for the total cross sections and $K$-factors obtained in the benchmark scenarios $A_{1,3}$ and $B_{2,3}$.

\begin{table}[h]
\small\centering
\begin{tabular}{|c|cccccccccc|}
\hline
 & \multicolumn{10}{|c|}{$M_{\textrm{med}}$\,[GeV]} \\
\cline{2-11}
 & 10 & 20 & 50 & 100 & 200 & 300 & 500 & 1000 & 2000 & 10000 \\
$m_{\textrm{DM}}$\,[GeV] & & (15) & & (95) & & (295) & & (995) & (1995) & \\
\hline
1 & 8.1 & 3.6 & 1.0 & 0.35 & 0.10 & 4.4e-2 & 1.3e-2 & 1.7e-3 & 1.1e-4 & 1.3e-8 \\
10 & 8.8e-2 & 0.13 & 1.0 & 0.34 & & & & & & 1.3e-8 \\
50 & 5.1e-3 & & 6.4e-3 & 2.9e-2 & 0.10 & 4.4e-2 & & & & 1.3e-8 \\
150 & 4.7e-4 & & & & 8.2e-4 & 6.6e-3 & 1.2e-2 & 1.7e-3 & & 1.2e-8 \\
500 & 1.1e-5 & & & & & & 1.5e-5 & 3.5e-4 & 1.0e-4 & 6.9e-9 \\
1000 & 4.1e-7 & & & & & & & 6.1e-7 & 2.1e-5 & 2.4e-9 \\
\hline
$m_{\textrm{DM}}$\,[GeV] & \multicolumn{10}{|c|}{$K$-factor} \\
\hline
1 & 1.52 & 1.52 & 1.52 & 1.47 & 1.43 & 1.39 & 1.38 & 1.36 & 1.30 & 1.30 \\
10 & 1.51 & 1.50 & 1.51 & 1.45 & & & & & & 1.28 \\
50 & 1.43 & & 1.44 & 1.46 & 1.42 & 1.39 & & & & 1.31 \\
150 & 1.38 & & & & 1.38 & 1.41 & 1.38 & 1.36 & & 1.26 \\
500 & 1.35 & & & & & & 1.36 & 1.37 & 1.31 & 1.26 \\
1000 & 1.26 & & & & & & & 1.27 & 1.31 & 1.16 \\
\hline
\end{tabular}
\caption{NLO total cross sections (in pb) and $K$-factors for mono-$Z$ production in the channel $pp\to Z(\to\mu^+\mu^-)+\cancel{E}_T$ at the 13 TeV LHC in benchmark scenario $A_1$. We use the short-hand notation ``e-$n$'' for $10^{-n}$. The values of $M_{\textrm{med}}$ shown in parenthesis are used at threshold $M_{\textrm{med}}=2m_{\textrm{DM}}$.}
\label{tab:nlo1}
\end{table}

\begin{table}[h]
\small\centering
\begin{tabular}{|c|cccccccccc|}
\hline
 & \multicolumn{10}{|c|}{$M_{\textrm{med}}$\,[GeV]} \\
\cline{2-11}
 & 10 & 20 & 50 & 100 & 200 & 300 & 500 & 1000 & 2000 & 10000 \\
$m_{\textrm{DM}}$\,[GeV] & & (15) & & (95) & & (295) & & (995) & (1995) & \\
\hline
1 & 0.98 & 0.44 & 0.13 & 4.2e-2 & 1.2e-2 & 5.5e-3 & 1.4e-3 & 1.9e-4 & 1.2e-5 & 7.9e-10 \\
10 & 2.9e-3 & 3.6e-3 & 0.10 & 4.1e-2 & & & & & & 7.8e-10 \\
50 & 1.6e-4 & & 1.9e-4 &4.1e-4 & 8.2e-3 & 4.7e-3 & & & & 7.5e-10 \\
150 & 1.3e-5 & & & & 1.9e-5 & 5.3e-5 & 7.4e-4 & 1.6e-4 & & 6.3e-10 \\
500 & 2.2e-7 & & & & & & 2.8e-7 & 1.8e-6 & 7.4e-6 & 2.6e-10 \\
1000 & 6.2e-9 & & & & & & & 8.5e-9 & 8.3e-8 & 6.0e-11 \\
\hline
$m_{\textrm{DM}}$\,[GeV] & \multicolumn{10}{|c|}{$K$-factor} \\
\hline
1 & 1.54 & 1.54 & 1.52 & 1.48 & 1.42 & 1.38 & 1.39 & 1.37 & 1.32 & 1.32 \\
10 & 1.48 & 1.49 & 1.52 & 1.49 & & & & & & 1.33 \\
50 & 1.41 & & 1.43 & 1.42 & 1.43 & 1.42 & & & & 1.28 \\
150 & 1.38 & & & & 1.39 & 1.38 & 1.37 & 1.37 & & 1.29 \\
500 & 1.35 & & & & & & 1.34 & 1.37 & 1.32 & 1.23 \\
1000 & 1.23 & & & & & & & 1.24 & 1.29 & 1.13 \\
\hline
\end{tabular}
\caption{NLO total cross sections (in pb) and $K$-factors for mono-$Z$ production in the channel $pp\to Z(\to\mu^+\mu^-)+\cancel{E}_T$ at the 13 TeV LHC in benchmark scenario $A_3$. We use the short-hand notation ``e-$n$'' for $10^{-n}$. The values of $M_{\textrm{med}}$ shown in parenthesis are used at threshold $M_{\textrm{med}}=2m_{\textrm{DM}}$.}
  \label{tab:nlo3}
\end{table}

\begin{table}[h]
\small\centering
\begin{tabular}{|c|cccccccccc|}
\hline
 & \multicolumn{10}{|c|}{$M_{\textrm{med}}$\,[GeV]} \\
\cline{2-11}
 & 10 & 20 & 50 & 100 & 200 & 300 & 500 & 1000 & 2000 & 10000 \\
$m_{\textrm{DM}}$\,[GeV] & & (15) & & (95) & & (295) & & (995) & (1995) & \\
\hline
1 & 2.0e-3 & 1.7e-3 & 9.2e-4 & 5.3e-4 & 2.4e-4 & 1.2e-4 & 4.3e-5 & 6.2e-6 & 3.0e-7 & 2.8e-11 \\
10 & 4.8e-5 & 6.2e-5 & 9.2e-4 & 5.3e-4 & & & & & & 2.8e-11 \\
50 & 7.1e-6 & & 8.6e-6 & 3.0e-5 & 2.3e-4 & 1.2e-4 & & & & 2.8e-11 \\
150 & 9.2e-7 & & & & 1.5e-6 & 1.1e-5 & 4.3e-5 & 6.1e-6 & & 2.6e-11 \\
500 & 2.2e-8 & & & & & & 3.1e-8 & 1.0e-6 & 2.8e-7 & 1.2e-11 \\
1000 & 2.2e-8 & & & & & & & 4.3e-6 & 2.8e-7 & 1.1e-11 \\
\hline
$m_{\textrm{DM}}$\,[GeV] & \multicolumn{10}{|c|}{$K$-factor} \\
\hline
1 & 1.38 & 1.39 & 1.35 & 1.33 & 1.32 & 1.31 & 1.31 & 1.30 & 1.16 & 1.23 \\
10 & 1.34 & 1.34 & 1.34 & 1.33 & & & & & & 1.20 \\
50 & 1.31 & & 1.31 & 1.33 & 1.31 & 1.31 & & & & 1.21 \\
150 & 1.31 & & & & 1.31 & 1.32 & 1.31 & 1.27 & & 1.23 \\
500 & 1.25 & & & & & & 1.25 & 1.28 & 1.18 & 1.12 \\
1000 & 1.24 & & & & & & & 1.31 & 1.18 & 1.00 \\
\hline
\end{tabular}
\caption{NLO total cross sections (in pb) and $K$-factors for mono-$Z$ production in the channel $pp\to Z(\to\mu^+\mu^-)+\cancel{E}_T$ at the 13 TeV LHC in benchmark scenario $B_2$. We use the short-hand notation ``e-$n$'' for $10^{-n}$. The values of $M_{\textrm{med}}$ shown in parenthesis are used at threshold $M_{\textrm{med}}=2m_{\textrm{DM}}$.}
\label{tab:nlo5}
\end{table}

\begin{table}[h]
\small\centering
\begin{tabular}{|c|cccccccccc|}
\hline
 & \multicolumn{10}{|c|}{$M_{\textrm{med}}$\,[GeV]} \\
\cline{2-11}
 & 10 & 20 & 50 & 100 & 200 & 300 & 500 & 1000 & 2000 & 10000 \\
$m_{\textrm{DM}}$\,[GeV] & & (15) & & (95) & & (295) & & (995) & (1995) & \\
\hline
1 & 1.2e-5 & 2.5e-6 & 1.2e-7 & 9.1e-9 & 5.0e-10 & 7.6e-11 & 6.0e-12 & 1.5e-13 & 3.0e-15 & 3.3e-18 \\
10 & 1.5e-6 & 2.7e-6 & 1.1e-5 & 8.7e-7 & & & & & & 1.7e-16 \\
50 & 1.4e-8 & & 1.9e-8 & 1.6e-7 & 1.0e-6 & 1.7e-7 & & & & 7.3e-16 \\
150 & 2.6e-10 & & & & 5.6e-10 & 8.5e-9 & 9.9e-8 & 2.4e-9 & & 8.1e-16 \\
500 & 9.8e-13 & & & & & & 1.5e-12 & 9.5e-11 & 1.9e-10 & 2.6e-16 \\
1000 & 9.8e-13 & & & & & & & 5.6e-10 & 1.9e-10 & 2.6e-16 \\
\hline
$m_{\textrm{DM}}$\,[GeV] & \multicolumn{10}{|c|}{$K$-factor}  \\
\hline
1 & 1.48 & 1.47 & 1.41 & 1.36 & 1.33 & 1.34 & 1.33 & 1.39 & 1.37 & 1.42 \\
10 & 1.43 & 1.43 & 1.42 & 1.35 & & & & & & 1.40 \\
50 & 1.36 & & 1.35 & 1.32 & 1.33 & 1.33 & & & & 1.36 \\
150 & 1.33 & & & & 1.33 & 1.32 & 1.31 & 1.31 & & 1.30 \\
500 & 1.32 & & & & & & 1.31 & 1.33 & 1.20 & 1.30 \\
1000 & 1.31 & & & & & & & 1.32 & 1.21 & 1.27 \\
\hline
\end{tabular}
\caption{NLO total cross sections (in pb) and $K$-factors for mono-$Z$ production in the channel $pp\to Z(\to\mu^+\mu^-)+\cancel{E}_T$ at the 13 TeV LHC in benchmark scenario $B_3$. We use the short-hand notation ``e-$n$'' for $10^{-n}$. The values of $M_{\textrm{med}}$ shown in parenthesis are used at threshold $M_{\textrm{med}}=2m_{\textrm{DM}}$.}
\label{tab:nlo6}
\end{table}

\bibliography{DMnlo}

\providecommand{\href}[2]{#2}\begingroup\raggedright\begin{thebibliography}{10}

\bibitem{Baer:2014eja}
H.~Baer, K.-Y. Choi, J.~E. Kim, and L.~Roszkowski, {\it {Dark matter production
  in the early Universe: beyond the thermal WIMP paradigm}},  {\em Phys.Rept.}
  {\bf 555} (2014) 1--60, [\href{http://arxiv.org/abs/1407.0017}{{\tt
  arXiv:1407.0017}}].

\bibitem{Gelmini:2015zpa}
G.~B. Gelmini, {\it {TASI 2014 Lectures: The Hunt for Dark Matter}},
  \href{http://arxiv.org/abs/1502.01320}{{\tt arXiv:1502.01320}}.

\bibitem{ATLAS:2012zim}
{\bf ATLAS} Collaboration, {\it {Search for New Phenomena in Monojet plus
  Missing Transverse Momentum Final States using 10fb-1 of pp Collisions at
  sqrt{s}=8 TeV with the ATLAS detector at the LHC}}, .

\bibitem{Aad:2012fw}
{\bf ATLAS} Collaboration, G.~Aad et~al., {\it {Search for dark matter
  candidates and large extra dimensions in events with a photon and missing
  transverse momentum in $pp$ collision data at $\sqrt{s}=7$ TeV with the ATLAS
  detector}},  {\em Phys.Rev.Lett.} {\bf 110} (2013), no.~1 011802,
  [\href{http://arxiv.org/abs/1209.4625}{{\tt arXiv:1209.4625}}].

\bibitem{Aad:2013oja}
{\bf ATLAS} Collaboration, G.~Aad et~al., {\it {Search for dark matter in
  events with a hadronically decaying W or Z boson and missing transverse
  momentum in $pp$ collisions at $\sqrt{s} =$ 8 TeV with the ATLAS detector}},
  {\em Phys.Rev.Lett.} {\bf 112} (2014), no.~4 041802,
  [\href{http://arxiv.org/abs/1309.4017}{{\tt arXiv:1309.4017}}].

\bibitem{Aad:2014vka}
{\bf ATLAS} Collaboration, G.~Aad et~al., {\it {Search for dark matter in
  events with a Z boson and missing transverse momentum in pp collisions at
  $\sqrt{s}$=8 TeV with the ATLAS detector}},  {\em Phys.Rev.} {\bf D90}
  (2014), no.~1 012004, [\href{http://arxiv.org/abs/1404.0051}{{\tt
  arXiv:1404.0051}}].

\bibitem{Khachatryan:2014rra}
{\bf CMS} Collaboration, V.~Khachatryan et~al., {\it {Search for dark matter,
  extra dimensions, and unparticles in monojet events in proton¨Cproton
  collisions at $\sqrt{s} = 8$ TeV}},  {\em Eur. Phys. J.} {\bf C75} (2015),
  no.~5 235, [\href{http://arxiv.org/abs/1408.3583}{{\tt arXiv:1408.3583}}].

\bibitem{Khachatryan:2014rwa}
{\bf CMS} Collaboration, V.~Khachatryan et~al., {\it {Search for new phenomena
  in monophoton final states in proton-proton collisions at $\sqrt{s}$ = 8
  TeV}},  \href{http://arxiv.org/abs/1410.8812}{{\tt arXiv:1410.8812}}.

\bibitem{Akerib:2013tjd}
{\bf LUX} Collaboration, D.~Akerib et~al., {\it {First results from the LUX
  dark matter experiment at the Sanford Underground Research Facility}},  {\em
  Phys.Rev.Lett.} {\bf 112} (2014) 091303,
  [\href{http://arxiv.org/abs/1310.8214}{{\tt arXiv:1310.8214}}].

\bibitem{Cao:2009uw}
Q.-H. Cao, C.-R. Chen, C.~S. Li, and H.~Zhang, {\it {Effective Dark Matter
  Model: Relic density, CDMS II, Fermi LAT and LHC}},  {\em JHEP} {\bf 1108}
  (2011) 018, [\href{http://arxiv.org/abs/0912.4511}{{\tt arXiv:0912.4511}}].

\bibitem{Bai:2010hh}
Y.~Bai, P.~J. Fox, and R.~Harnik, {\it {The Tevatron at the Frontier of Dark
  Matter Direct Detection}},  {\em JHEP} {\bf 1012} (2010) 048,
  [\href{http://arxiv.org/abs/1005.3797}{{\tt arXiv:1005.3797}}].

\bibitem{Goodman:2010ku}
J.~Goodman, M.~Ibe, A.~Rajaraman, W.~Shepherd, T.~M. Tait, et~al., {\it
  {Constraints on Dark Matter from Colliders}},  {\em Phys.Rev.} {\bf D82}
  (2010) 116010, [\href{http://arxiv.org/abs/1008.1783}{{\tt
  arXiv:1008.1783}}].

\bibitem{Beltran:2010ww}
M.~Beltran, D.~Hooper, E.~W. Kolb, Z.~A. Krusberg, and T.~M. Tait, {\it
  {Maverick dark matter at colliders}},  {\em JHEP} {\bf 1009} (2010) 037,
  [\href{http://arxiv.org/abs/1002.4137}{{\tt arXiv:1002.4137}}].

\bibitem{Fox:2011pm}
P.~J. Fox, R.~Harnik, J.~Kopp, and Y.~Tsai, {\it {Missing Energy Signatures of
  Dark Matter at the LHC}},  {\em Phys.Rev.} {\bf D85} (2012) 056011,
  [\href{http://arxiv.org/abs/1109.4398}{{\tt arXiv:1109.4398}}].

\bibitem{Shoemaker:2011vi}
I.~M. Shoemaker and L.~Vecchi, {\it {Unitarity and Monojet Bounds on Models for
  DAMA, CoGeNT, and CRESST-II}},  {\em Phys. Rev.} {\bf D86} (2012) 015023,
  [\href{http://arxiv.org/abs/1112.5457}{{\tt arXiv:1112.5457}}].

\bibitem{Busoni:2013lha}
G.~Busoni, A.~De~Simone, E.~Morgante, and A.~Riotto, {\it {On the Validity of
  the Effective Field Theory for Dark Matter Searches at the LHC}},  {\em
  Phys.Lett.} {\bf B728} (2014) 412--421,
  [\href{http://arxiv.org/abs/1307.2253}{{\tt arXiv:1307.2253}}].

\bibitem{Buchmueller:2013dya}
O.~Buchmueller, M.~J. Dolan, and C.~McCabe, {\it {Beyond Effective Field Theory
  for Dark Matter Searches at the LHC}},  {\em JHEP} {\bf 1401} (2014) 025,
  [\href{http://arxiv.org/abs/1308.6799}{{\tt arXiv:1308.6799}}].

\bibitem{Busoni:2014sya}
G.~Busoni, A.~De~Simone, J.~Gramling, E.~Morgante, and A.~Riotto, {\it {On the
  Validity of the Effective Field Theory for Dark Matter Searches at the LHC,
  Part II: Complete Analysis for the $s$-channel}},  {\em JCAP} {\bf 1406}
  (2014) 060, [\href{http://arxiv.org/abs/1402.1275}{{\tt arXiv:1402.1275}}].

\bibitem{Busoni:2014haa}
G.~Busoni, A.~De~Simone, T.~Jacques, E.~Morgante, and A.~Riotto, {\it {On the
  Validity of the Effective Field Theory for Dark Matter Searches at the LHC
  Part III: Analysis for the $t$-channel}},  {\em JCAP} {\bf 1409} (2014) 022,
  [\href{http://arxiv.org/abs/1405.3101}{{\tt arXiv:1405.3101}}].

\bibitem{Alwall:2008ag}
J.~Alwall, P.~Schuster, and N.~Toro, {\it {Simplified Models for a First
  Characterization of New Physics at the LHC}},  {\em Phys.Rev.} {\bf D79}
  (2009) 075020, [\href{http://arxiv.org/abs/0810.3921}{{\tt
  arXiv:0810.3921}}].

\bibitem{Alves:2011wf}
{\bf LHC New Physics Working Group} Collaboration, D.~Alves et~al., {\it
  {Simplified Models for LHC New Physics Searches}},  {\em J.Phys.} {\bf G39}
  (2012) 105005, [\href{http://arxiv.org/abs/1105.2838}{{\tt
  arXiv:1105.2838}}].

\bibitem{Goodman:2011jq}
J.~Goodman and W.~Shepherd, {\it {LHC Bounds on UV-Complete Models of Dark
  Matter}},  \href{http://arxiv.org/abs/1111.2359}{{\tt arXiv:1111.2359}}.

\bibitem{An:2012va}
H.~An, X.~Ji, and L.-T. Wang, {\it {Light Dark Matter and $Z'$ Dark Force at
  Colliders}},  {\em JHEP} {\bf 1207} (2012) 182,
  [\href{http://arxiv.org/abs/1202.2894}{{\tt arXiv:1202.2894}}].

\bibitem{Frandsen:2012rk}
M.~T. Frandsen, F.~Kahlhoefer, A.~Preston, S.~Sarkar, and K.~Schmidt-Hoberg,
  {\it {LHC and Tevatron Bounds on the Dark Matter Direct Detection
  Cross-Section for Vector Mediators}},  {\em JHEP} {\bf 1207} (2012) 123,
  [\href{http://arxiv.org/abs/1204.3839}{{\tt arXiv:1204.3839}}].

\bibitem{An:2013xka}
H.~An, L.-T. Wang, and H.~Zhang, {\it {Dark matter with $t$-channel mediator: a
  simple step beyond contact interaction}},  {\em Phys.Rev.} {\bf D89} (2014),
  no.~11 115014, [\href{http://arxiv.org/abs/1308.0592}{{\tt
  arXiv:1308.0592}}].

\bibitem{DiFranzo:2013vra}
A.~DiFranzo, K.~I. Nagao, A.~Rajaraman, and T.~M. Tait, {\it {Simplified Models
  for Dark Matter Interacting with Quarks}},  {\em JHEP} {\bf 1311} (2013) 014,
  [\href{http://arxiv.org/abs/1308.2679}{{\tt arXiv:1308.2679}}].

\bibitem{Papucci:2014iwa}
M.~Papucci, A.~Vichi, and K.~M. Zurek, {\it {Monojet versus the rest of the
  world I: t-channel models}},  {\em JHEP} {\bf 1411} (2014) 024,
  [\href{http://arxiv.org/abs/1402.2285}{{\tt arXiv:1402.2285}}].

\bibitem{Berlin:2014tja}
A.~Berlin, D.~Hooper, and S.~D. McDermott, {\it {Simplified Dark Matter Models
  for the Galactic Center Gamma-Ray Excess}},  {\em Phys.Rev.} {\bf D89}
  (2014), no.~11 115022, [\href{http://arxiv.org/abs/1404.0022}{{\tt
  arXiv:1404.0022}}].

\bibitem{Buchmueller:2014yoa}
O.~Buchmueller, M.~J. Dolan, S.~A. Malik, and C.~McCabe, {\it {Characterising
  dark matter searches at colliders and direct detection experiments: Vector
  mediators}},  {\em JHEP} {\bf 1501} (2015) 037,
  [\href{http://arxiv.org/abs/1407.8257}{{\tt arXiv:1407.8257}}].

\bibitem{Abdallah:2014hon}
J.~Abdallah, A.~Ashkenazi, A.~Boveia, G.~Busoni, A.~De~Simone, et~al., {\it
  {Simplified Models for Dark Matter and Missing Energy Searches at the LHC}},
  \href{http://arxiv.org/abs/1409.2893}{{\tt arXiv:1409.2893}}.

\bibitem{Malik:2014ggr}
S.~Malik, C.~McCabe, H.~Araujo, A.~Belyaev, C.~Boehm, et~al., {\it {Interplay
  and Characterization of Dark Matter Searches at Colliders and in Direct
  Detection Experiments}},  \href{http://arxiv.org/abs/1409.4075}{{\tt
  arXiv:1409.4075}}.

\bibitem{Buckley:2014fba}
M.~R. Buckley, D.~Feld, and D.~Goncalves, {\it {Scalar Simplified Models for
  Dark Matter}},  {\em Phys.Rev.} {\bf D91} (2015), no.~1 015017,
  [\href{http://arxiv.org/abs/1410.6497}{{\tt arXiv:1410.6497}}].

\bibitem{Harris:2014hga}
P.~Harris, V.~V. Khoze, M.~Spannowsky, and C.~Williams, {\it {Constraining Dark
  Sectors at Colliders: Beyond the Effective Theory Approach}},  {\em Phys.
  Rev.} {\bf D91} (2015) 055009, [\href{http://arxiv.org/abs/1411.0535}{{\tt
  arXiv:1411.0535}}].

\bibitem{Alves:2015pea}
A.~Alves, A.~Berlin, S.~Profumo, and F.~S. Queiroz, {\it {Dark Matter
  Complementarity and the Z$^\prime$ Portal}},
  \href{http://arxiv.org/abs/1501.03490}{{\tt arXiv:1501.03490}}.

\bibitem{Jacques:2015zha}
T.~Jacques and K.~Nordstrom, {\it {Mapping monojet constraints onto Simplified
  Dark Matter Models}},  {\em JHEP} {\bf 06} (2015) 142,
  [\href{http://arxiv.org/abs/1502.05721}{{\tt arXiv:1502.05721}}].

\bibitem{Haisch:2015ioa}
U.~Haisch and E.~Re, {\it {Simplified dark matter top-quark interactions at the
  LHC}},  {\em JHEP} {\bf 06} (2015) 078,
  [\href{http://arxiv.org/abs/1503.00691}{{\tt arXiv:1503.00691}}].

\bibitem{Alves:2015mua}
A.~Alves, A.~Berlin, S.~Profumo, and F.~S. Queiroz, {\it {Dirac-Fermionic Dark
  Matter in $U(1)_X$ Models}},  \href{http://arxiv.org/abs/1506.06767}{{\tt
  arXiv:1506.06767}}.

\bibitem{Harris:2015kda}
P.~Harris, V.~V. Khoze, M.~Spannowsky, and C.~Williams, {\it {Closing up on
  Dark Sectors at Colliders: from 14 to 100 TeV}},
  \href{http://arxiv.org/abs/1509.02904}{{\tt arXiv:1509.02904}}.

\bibitem{Abdallah:2015ter}
J.~Abdallah et~al., {\it {Simplified Models for Dark Matter Searches at the
  LHC}},  \href{http://arxiv.org/abs/1506.03116}{{\tt arXiv:1506.03116}}.

\bibitem{Abercrombie:2015wmb}
D.~Abercrombie et~al., {\it {Dark Matter Benchmark Models for Early LHC Run-2
  Searches: Report of the ATLAS/CMS Dark Matter Forum}},
  \href{http://arxiv.org/abs/1507.00966}{{\tt arXiv:1507.00966}}.

\bibitem{Wang:2011sx}
J.~Wang, C.~S. Li, D.~Y. Shao, and H.~Zhang, {\it {Next-to-leading order QCD
  predictions for the signal of Dark Matter and photon associated production at
  the LHC}},  {\em Phys.Rev.} {\bf D84} (2011) 075011,
  [\href{http://arxiv.org/abs/1107.2048}{{\tt arXiv:1107.2048}}].

\bibitem{Huang:2012hs}
F.~P. Huang, C.~S. Li, J.~Wang, and D.~Y. Shao, {\it {Searching for the signal
  of dark matter and photon associated production at the LHC beyond leading
  order}},  {\em Phys.Rev.} {\bf D87} (2013) 094018,
  [\href{http://arxiv.org/abs/1210.0195}{{\tt arXiv:1210.0195}}].

\bibitem{Fox:2012ru}
P.~J. Fox and C.~Williams, {\it {Next-to-Leading Order Predictions for Dark
  Matter Production at Hadron Colliders}},  {\em Phys.Rev.} {\bf D87} (2013)
  054030, [\href{http://arxiv.org/abs/1211.6390}{{\tt arXiv:1211.6390}}].

\bibitem{Haisch:2012kf}
U.~Haisch, F.~Kahlhoefer, and J.~Unwin, {\it {The impact of heavy-quark loops
  on LHC dark matter searches}},  {\em JHEP} {\bf 07} (2013) 125,
  [\href{http://arxiv.org/abs/1208.4605}{{\tt arXiv:1208.4605}}].

\bibitem{Haisch:2013ata}
U.~Haisch, F.~Kahlhoefer, and E.~Re, {\it {QCD effects in mono-jet searches for
  dark matter}},  {\em JHEP} {\bf 1312} (2013) 007,
  [\href{http://arxiv.org/abs/1310.4491}{{\tt arXiv:1310.4491}}].

\bibitem{Mao:2014rga}
M.~Song, G.~Li, W.-G. Ma, R.-Y. Zhang, and J.-Y. Guo, {\it {Dark matter pair
  associated with a $W$ boson production at the LHC in next-to-leading order
  QCD}},  {\em JHEP} {\bf 09} (2014) 069,
  [\href{http://arxiv.org/abs/1403.2142}{{\tt arXiv:1403.2142}}].

\bibitem{Carpenter:2012rg}
L.~M. Carpenter, A.~Nelson, C.~Shimmin, T.~M.~P. Tait, and D.~Whiteson, {\it
  {Collider searches for dark matter in events with a Z boson and missing
  energy}},  {\em Phys. Rev.} {\bf D87} (2013), no.~7 074005,
  [\href{http://arxiv.org/abs/1212.3352}{{\tt arXiv:1212.3352}}].

\bibitem{Bell:2012rg}
N.~F. Bell, J.~B. Dent, A.~J. Galea, T.~D. Jacques, L.~M. Krauss, et~al., {\it
  {Searching for Dark Matter at the LHC with a Mono-Z}},  {\em Phys.Rev.} {\bf
  D86} (2012) 096011, [\href{http://arxiv.org/abs/1209.0231}{{\tt
  arXiv:1209.0231}}].

\bibitem{Chen:2013gya}
J.-Y. Chen, E.~W. Kolb, and L.-T. Wang, {\it {Dark matter coupling to
  electroweak gauge and Higgs bosons: an effective field theory approach}},
  {\em Phys.Dark Univ.} {\bf 2} (2013) 200--218,
  [\href{http://arxiv.org/abs/1305.0021}{{\tt arXiv:1305.0021}}].

\bibitem{Alves:2015dya}
A.~Alves and K.~Sinha, {\it {Searches for Dark Matter at the LHC: A
  Multivariate Analysis in the Mono-$Z$ Channel}},
  \href{http://arxiv.org/abs/1507.08294}{{\tt arXiv:1507.08294}}.

\bibitem{Crivellin:2015wva}
A.~Crivellin, U.~Haisch, and A.~Hibbs, {\it {LHC constraints on gauge boson
  couplings to dark matter}},  {\em Phys. Rev.} {\bf D91} (2015) 074028,
  [\href{http://arxiv.org/abs/1501.00907}{{\tt arXiv:1501.00907}}].

\bibitem{Aad:2012awa}
{\bf ATLAS} Collaboration, G.~Aad et~al., {\it {Measurement of $ZZ$ production
  in $pp$ collisions at $\sqrt{s}=7$ TeV and limits on anomalous $ZZZ$ and
  $ZZ\gamma$ couplings with the ATLAS detector}},  {\em JHEP} {\bf 1303} (2013)
  128, [\href{http://arxiv.org/abs/1211.6096}{{\tt arXiv:1211.6096}}].

\bibitem{Alloul:2013bka}
A.~Alloul, N.~D. Christensen, C.~Degrande, C.~Duhr, and B.~Fuks, {\it
  {FeynRules 2.0 - A complete toolbox for tree-level phenomenology}},  {\em
  Comput. Phys. Commun.} {\bf 185} (2014) 2250--2300,
  [\href{http://arxiv.org/abs/1310.1921}{{\tt arXiv:1310.1921}}].

\bibitem{Alwall:2014hca}
J.~Alwall, R.~Frederix, S.~Frixione, V.~Hirschi, F.~Maltoni, et~al., {\it {The
  automated computation of tree-level and next-to-leading order differential
  cross sections, and their matching to parton shower simulations}},  {\em
  JHEP} {\bf 1407} (2014) 079, [\href{http://arxiv.org/abs/1405.0301}{{\tt
  arXiv:1405.0301}}].

\bibitem{Backovic:2015soa}
M.~Backovic, M.~Kramer, F.~Maltoni, A.~Martini, K.~Mawatari, and M.~Pellen,
  {\it {Higher-order QCD predictions for dark matter production at the LHC in
  simplified models with s-channel mediators}},
  \href{http://arxiv.org/abs/1508.05327}{{\tt arXiv:1508.05327}}.

\bibitem{Mattelaer:2015haa}
O.~Mattelaer and E.~Vryonidou, {\it {Dark matter production through
  loop-induced processes at the LHC: the s-channel mediator case}},  {\em Eur.
  Phys. J.} {\bf C75} (2015), no.~9 436,
  [\href{http://arxiv.org/abs/1508.00564}{{\tt arXiv:1508.00564}}].

\bibitem{Hirschi:2015iia}
V.~Hirschi and O.~Mattelaer, {\it {Automated event generation for loop-induced
  processes}},  \href{http://arxiv.org/abs/1507.00020}{{\tt arXiv:1507.00020}}.

\bibitem{dmsimp}
\url{http://feynrules.irmp.ucl.ac.be/wiki/DMsimp}.

\bibitem{Cheung:2010zf}
K.~Cheung, K.~Mawatari, E.~Senaha, P.-Y. Tseng, and T.-C. Yuan, {\it {The Top
  Window for dark matter}},  {\em JHEP} {\bf 10} (2010) 081,
  [\href{http://arxiv.org/abs/1009.0618}{{\tt arXiv:1009.0618}}].

\bibitem{Lin:2013sca}
T.~Lin, E.~W. Kolb, and L.-T. Wang, {\it {Probing dark matter couplings to top
  and bottom at the LHC}},  {\em Phys.Rev.} {\bf D88} (2013) 063510,
  [\href{http://arxiv.org/abs/1303.6638}{{\tt arXiv:1303.6638}}].

\bibitem{Agashe:2014kda}
{\bf Particle Data Group} Collaboration, K.~A. Olive et~al., {\it {Review of
  Particle Physics}},  {\em Chin. Phys.} {\bf C38} (2014) 090001.

\bibitem{Khachatryan:2014jba}
{\bf CMS} Collaboration, V.~Khachatryan et~al., {\it {Precise determination of
  the mass of the Higgs boson and tests of compatibility of its couplings with
  the standard model predictions using proton collisions at 7 and 8 $\,\text
  {TeV}$}},  {\em Eur. Phys. J.} {\bf C75} (2015), no.~5 212,
  [\href{http://arxiv.org/abs/1412.8662}{{\tt arXiv:1412.8662}}].

\bibitem{Baek:2011aa}
S.~Baek, P.~Ko, and W.-I. Park, {\it {Search for the Higgs portal to a singlet
  fermionic dark matter at the LHC}},  {\em JHEP} {\bf 02} (2012) 047,
  [\href{http://arxiv.org/abs/1112.1847}{{\tt arXiv:1112.1847}}].

\bibitem{LopezHonorez:2012kv}
L.~Lopez-Honorez, T.~Schwetz, and J.~Zupan, {\it {Higgs portal, fermionic dark
  matter, and a Standard Model like Higgs at 125 GeV}},  {\em Phys. Lett.} {\bf
  B716} (2012) 179--185, [\href{http://arxiv.org/abs/1203.2064}{{\tt
  arXiv:1203.2064}}].

\bibitem{Baek:2012se}
S.~Baek, P.~Ko, W.-I. Park, and E.~Senaha, {\it {Higgs Portal Vector Dark
  Matter : Revisited}},  {\em JHEP} {\bf 05} (2013) 036,
  [\href{http://arxiv.org/abs/1212.2131}{{\tt arXiv:1212.2131}}].

\bibitem{Duch:2015jta}
M.~Duch, B.~Grzadkowski, and M.~McGarrie, {\it {A stable Higgs portal with
  vector dark matter}},  \href{http://arxiv.org/abs/1506.08805}{{\tt
  arXiv:1506.08805}}.

\bibitem{Degrande:2011ua}
C.~Degrande, C.~Duhr, B.~Fuks, D.~Grellscheid, O.~Mattelaer, and T.~Reiter,
  {\it {UFO - The Universal FeynRules Output}},  {\em Comput. Phys. Commun.}
  {\bf 183} (2012) 1201--1214, [\href{http://arxiv.org/abs/1108.2040}{{\tt
  arXiv:1108.2040}}].

\bibitem{Cotta:2012nj}
R.~C. Cotta, J.~L. Hewett, M.~P. Le, and T.~G. Rizzo, {\it {Bounds on Dark
  Matter Interactions with Electroweak Gauge Bosons}},  {\em Phys. Rev.} {\bf
  D88} (2013) 116009, [\href{http://arxiv.org/abs/1210.0525}{{\tt
  arXiv:1210.0525}}].

\bibitem{Hirschi:2011pa}
V.~Hirschi, R.~Frederix, S.~Frixione, M.~V. Garzelli, F.~Maltoni, and
  R.~Pittau, {\it {Automation of one-loop QCD corrections}},  {\em JHEP} {\bf
  05} (2011) 044, [\href{http://arxiv.org/abs/1103.0621}{{\tt
  arXiv:1103.0621}}].

\bibitem{Ossola:2006us}
G.~Ossola, C.~G. Papadopoulos, and R.~Pittau, {\it {Reducing full one-loop
  amplitudes to scalar integrals at the integrand level}},  {\em Nucl. Phys.}
  {\bf B763} (2007) 147--169, [\href{http://arxiv.org/abs/hep-ph/0609007}{{\tt
  hep-ph/0609007}}].

\bibitem{Ossola:2007ax}
G.~Ossola, C.~G. Papadopoulos, and R.~Pittau, {\it {CutTools: A Program
  implementing the OPP reduction method to compute one-loop amplitudes}},  {\em
  JHEP} {\bf 03} (2008) 042, [\href{http://arxiv.org/abs/0711.3596}{{\tt
  arXiv:0711.3596}}].

\bibitem{Degrande:2014vpa}
C.~Degrande, {\it {Automatic evaluation of UV and R2 terms for beyond the
  Standard Model Lagrangians: a proof-of-principle}},
  \href{http://arxiv.org/abs/1406.3030}{{\tt arXiv:1406.3030}}.

\bibitem{Frederix:2009yq}
R.~Frederix, S.~Frixione, F.~Maltoni, and T.~Stelzer, {\it {Automation of
  next-to-leading order computations in QCD: The FKS subtraction}},  {\em JHEP}
  {\bf 10} (2009) 003, [\href{http://arxiv.org/abs/0908.4272}{{\tt
  arXiv:0908.4272}}].

\bibitem{Frixione:2002ik}
S.~Frixione and B.~R. Webber, {\it {Matching NLO QCD computations and parton
  shower simulations}},  {\em JHEP} {\bf 06} (2002) 029,
  [\href{http://arxiv.org/abs/hep-ph/0204244}{{\tt hep-ph/0204244}}].

\bibitem{Ball:2013hta}
{\bf NNPDF} Collaboration, R.~D. Ball, V.~Bertone, S.~Carrazza, L.~Del~Debbio,
  S.~Forte, A.~Guffanti, N.~P. Hartland, and J.~Rojo, {\it {Parton
  distributions with QED corrections}},  {\em Nucl. Phys.} {\bf B877} (2013)
  290--320, [\href{http://arxiv.org/abs/1308.0598}{{\tt arXiv:1308.0598}}].

\bibitem{Alwall:2014bza}
J.~Alwall, C.~Duhr, B.~Fuks, O.~Mattelaer, D.~G. Ozturk, and C.-H. Shen, {\it
  {Computing decay rates for new physics theories with FeynRules and
  MadGraph5/aMC@NLO}},  \href{http://arxiv.org/abs/1402.1178}{{\tt
  arXiv:1402.1178}}.

\bibitem{Sjostrand:2006za}
T.~Sjostrand, S.~Mrenna, and P.~Z. Skands, {\it {PYTHIA 6.4 Physics and
  Manual}},  {\em JHEP} {\bf 05} (2006) 026,
  [\href{http://arxiv.org/abs/hep-ph/0603175}{{\tt hep-ph/0603175}}].

\end{thebibliography}\endgroup
\bibliographystyle{JHEP}

\end{document}